\documentclass[twocolumn,tighten]{aastex62}

\usepackage{amsmath}
\usepackage{xcolor}

\newcommand{\keyw}[1]{\textcolor{gray}{#1}}

\newcommand{\dt}[1]{\textcolor{black}{#1}}
\begin{document}

\title{Scaling Relations Associated with Millimeter Continuum Sizes in Protoplanetary Disks}

\author{Sean M.~Andrews}
\affiliation{Harvard-Smithsonian Center for Astrophysics, 60 Garden Street, Cambridge, MA 02138, USA}
\author{Marie Terrell}
\affiliation{Harvard-Smithsonian Center for Astrophysics, 60 Garden Street, Cambridge, MA 02138, USA}
\affiliation{University of Massachusetts, Lowell, MA 01854, USA}
\author{Anjali Tripathi}
\affiliation{Harvard-Smithsonian Center for Astrophysics, 60 Garden Street, Cambridge, MA 02138, USA} 
\author{Megan Ansdell}
\affiliation{University of Hawaii Institute for Astronomy, 2680 Woodlawn Drive, Honolulu, HI 97822, USA}
\affiliation{Department of Astronomy, University of California, Berkeley, CA 94720, USA}
\affiliation{Center for Integrative Planetary Science, University of California, Berkeley, CA 94720, USA}
\author{Jonathan P.~Williams}
\affiliation{University of Hawaii Institute for Astronomy, 2680 Woodlawn Drive, Honolulu, HI 97822, USA}
\author{David J.~Wilner}
\affiliation{Harvard-Smithsonian Center for Astrophysics, 60 Garden Street, Cambridge, MA 02138, USA}

\begin{abstract}
We present a combined, homogenized analysis of archival Submillimeter Array (SMA) and Atacama Large Millimeter/submillimeter Array (ALMA) observations of the spatially resolved 340\,GHz (870\,$\mu$m) continuum emission from 105 nearby protoplanetary disks.  Building on the previous SMA survey, we infer surface brightness profiles using a simple model of the observed visibilities to derive the luminosities ($L_{\rm mm}$) and effective sizes ($R_{\rm eff}$) of the continuum emission.  With this sample, we confirm the shapes, normalizations, and dispersions for the strong correlations between $L_{\rm mm}$, $M_\ast$ (or $L_\ast$), and $\dot{M}_\ast$ found in previous studies.  Moreover, we verify the continuum size--luminosity relation determined from the SMA survey alone (extending to an order of magnitude lower $L_{\rm mm}$), demonstrating that the amount of emission scales linearly with the emitting surface area.  Moreover, we identify new,  although weaker, relationships between $R_{\rm eff}$ and the host and accretion properties, such that disks are larger around more massive hosts with higher accretion rates.  We explore these inter-related demographic properties with some highly simplified approximations.  These multi-dimensional relationships can be explained if the emission is optically thick with a filling factor of $\sim$0.3, or if the emission is optically thin and disks have roughly the same optical depth profile shapes and normalizations independent of host properties.  In both scenarios, we require the dust disk sizes to have a slightly sub-linear relationship with the host mass and a non-negligible dispersion ($\sim$0.2\,dex at a given $M_\ast$).               
\end{abstract}

\section{Introduction \label{sec:intro}}

The physical processes most relevant to the formation and early evolution of planetary systems in the disks orbiting young stars are remarkably complex and, at present, only weakly constrained by the data.  Nevertheless, the prospect that the most dominant of these processes will imprint deterministic signatures in the ``bulk" characteristics of the disk population serves as a strong motivation for demographics studies \citep[e.g.,][]{ida04,ida08b,mordasini09,mordasini12}.  

Early successes in such population-oriented research were manifested in basic evolutionary trends.  Surveys found that the infrared excess fractions \citep{haisch01,hernandez07,richert18} and accretion rates ($\dot{M_\ast}$; \citealt{muzerolle00,sicilia-aguilar05}) in different clusters decreased with stellar host age ($\tau_\ast$), indicating that gas and dust in the innermost regions of disks are (exponentially) depleted over a (half-life) timescale of a few Myr \citep{mamajek09,fedele10}.  Those same metrics also depend on the stellar host mass ($M_\ast$), such that (at a given age) warm dust is more depleted and gas accretes faster in the disks around more massive stars \citep[e.g.,][]{carpenter06,muzerolle03,manara17}.  Such behavior is broadly consistent with predictions for the viscous evolution of disk structures \citep{hartmann98}, but these specific diagnostics are not obviously associated with the more {\it global} disk properties that are most relevant for testing planet formation models.          

The global property of most interest is the disk mass ($M_d$).  By far, the most accessible estimates of $M_d$ for large-scale demographics studies come from the (presumed) optically thin millimeter-wavelength continuum emitted by dust grains in the disk \citep{beckwith90}.  Millimeter continuum luminosities ($L_{\rm mm}$) are expected to scale roughly linearly with $M_d$, modulo assumptions about the dust opacities ($\kappa_\nu$), dust-to-gas ratio, and temperature ($T_d$).  In population studies of several young ($\sim$1--3\,Myr) clusters, $L_{\rm mm}$ is found to have a steep dependence on $M_\ast$ that has been interpreted as a linear or super-linear scaling between $M_d$ and $M_\ast$ \citep{andrews13,ansdell16,pascucci16}.  Similar surveys in older clusters demonstrate that this relationship evolves (steepens) on few Myr timescales \citep{barenfeld16,ansdell17,ruiz-rodriguez18}.  Building on these results, \citet{manara16} also identified a tentative $M_d$ -- $\dot{M_\ast}$ relationship.     

Taken together, these studies affirm the promise of demographics to reveal fundamental links between the bulk properties of disks that are relevant to planet formation \citep[e.g.,][]{mulders17}.  They also confirm that the general disk population is rife with high-dimensionality connections: so far, there is evidence that dynamics and/or heating (via $M_\ast$), evolution ($\tau_\ast$), accretion ($\dot{M_\ast}$), and mass ($M_d$) are all related.  That said, the substantial amount of scatter around these confirmed relationships hints at other parameters of interest.    

Some of that scatter, and perhaps some of those relationships, might be associated with the ways in which mass is spatially distributed in these disks \citep[e.g.,][]{hartmann98,rafikov17}.  The relevant observable that is a complement to $L_{\rm mm}$ ($\sim$$M_d$) and suitable for demographics studies in that case is the spatial extent, or {\it size}, of the mm continuum emission.  \citet{andrews10b} first tentatively identified a positive trend between disk sizes and masses inferred from mm continuum data \citep[see also][]{pietu14}, which was robustly verified for a much larger sample by \citet{tripathi17}.  Recent work with other samples has broadly validated and extended this size--luminosity correlation \citep[albeit with somewhat different definitions of ``size";][]{tazzari17,barenfeld17}.  One challenge in linking the sizes with the other relevant parameters has been that the \citet{tripathi17} sample has a comparatively heterogeneous construction: it was primarily based on availability in the Submillimeter Array (SMA) archive, and thereby comprises smaller collections of disks in different cluster environments.                

In this article, we aim to alleviate some potential sample biases and re-examine the mm continuum size--luminosity relationship for a large and well-defined collection of disks in the Lupus star-forming region \citep{ansdell16}, with the goal of facilitating a more holistic look at its links to the larger disk demographics landscape.  Section~\ref{sec:data} describes the associated data, collected from the Atacama Large Millimeter/submillimeter Array (ALMA) archive.  Section~\ref{sec:sample} describes the sample.  Section~\ref{sec:analysis} explains the size measurements, and Section~\ref{sec:results} quantifies some relationships with other parameters.  Finally, Section~\ref{sec:discussion} considers potential origins for these connections in the context of the general disk population.

\section{Data \label{sec:data}}


The vast majority of the ALMA data used here were obtained on 2015 June 14 and 15 as part of Program 2013.1.00220.S.  The observations and calibration procedure were described in detail by \citet{ansdell16}.  To briefly summarize in the context of this article, this program employed the ALMA Band 7 receivers to observe the dust continuum emission at a mean frequency of 335\,GHz (895\,$\mu$m) with 41 and 37 of the 12\,m antennas arranged to span baselines of 21 to 784\,m.  The observations were ``snapshots", with total integration times of 48--120\,s per target.  The raw data were pipeline-calibrated by NRAO staff.  We then removed two narrowband spectral windows (containing spectral lines at high velocity resolution) and then spectrally averaged the data in 3 other spectral windows to 256\,MHz-wide pseudo-channels.  The total continuum bandwidth in those windows is $\sim$5\,GHz.  For brighter targets, a single iteration of phase-only self-calibration (on a 30\,s interval) was employed, which enhanced the peak S/N in the resulting emission maps by $\sim$20\%.  

Those data were supplemented with ALMA Band 7 observations of three additional targets -- IM Lup, Sz 91, and GQ Lup -- from Programs 2013.1.00694.S, 2012.1.00761.S, and 2013.1.00374.S, respectively.  These data and their calibration were already presented by \citet{cleeves16}, \citet{tsukagoshi17}, and \citet{macgregor17}, respectively.  IM Lup was observed on 2014 June 8 with 34 antennas configured to have baseline lengths of 28--646\,m.  The on-source integration time was $\sim$20 minutes.  After a single iteration of phase-only self-calibration, spectrally-averaged continuum visibilities were extracted from a $\sim$2\,GHz-wide window centered at 343\,GHz (875\,$\mu$m).  Sz 91 was observed on 2015 July 20 and 21, with 42 antennas distributed on baselines from 15 to 1574\,m, in two 1.875\,GHz-wide continuum windows.  The mean frequency for these data is 349\,GHz (858\,$\mu$m); the total on-source integration time was $\sim$90 minutes.  GQ Lup was observed on 2015 June 14 and 15, using 41 and 37 antennas, respectively, covering baseline lengths of 15 to 784\,m.  The combined integration time was about 30 minutes.  The visibilities were spectrally averaged after excising bright emission lines, providing a total continuum bandwidth of roughly 7\,GHz, and phase self-calibrated on 30\,s intervals. 


\section{Sample \label{sec:sample}}

The ALMA ``snapshot" survey of Lupus (Program 2013.1.00220.S; originally presented by \citealt{ansdell16}) includes 98 distinct pointings.  As noted by \citeauthor{ansdell16}, 8 of these pointings are toward targets that were later identified as background giants unrelated to the Lupus clouds (see their Table 4).  One additional pointing, toward Lupus MMS 3, was excluded because that target shows clear evidence for contamination by a circumstellar envelope \citep[e.g.,][]{yen15}.  Of the remaining 89 pointings, 24 have targets that are not detected.  An additional 3 are too faint ($<2$\,mJy) to be usefully included in our subsequent analysis.

The remaining 62 pointings actually contain 62 unique disks with continuum detections, albeit (confusingly) not in the sense of one detection per pointing.  Three pointings include close binary systems where both components are detected: V856 Sco, Sz 74, and Sz 81.  Two other systems, Sz 88 and Sz 123, are targeted in multiple overlapping pointings (2 and 3, respectively).   
 
In the analysis that follows, we have excluded the known multiple systems with projected separations $\lesssim 2$\arcsec.  That threshold is designed to remove disks that could have their \dt{continuum} sizes \dt{and luminosities} affected by dynamical interactions with their companions \citep[e.g.,][]{artymowicz94,jensen94,harris12,akeson14}.  The same criterion was adopted in the similar study by \citet{tripathi17}.  Accounting for these exclusions, the final sample includes 56 disks.  \dt{The Lupus population has not been thoroughly vetted for multiplicity, particularly at sub-arcsecond separations.  Presuming a similar binary fraction as in the Taurus Class II population \citep[e.g.,][]{kraus11}, we could be missing up to $\sim$6--10 additional multiple systems in this sample.  Due to their truncation, such interlopers may have already been excluded by the detection/flux density criterion noted above.}  

In Sections~\ref{sec:results} and \ref{sec:discussion}, we will fold in the \citet{tripathi17} results.  For clarity, when appropriate we will refer to the new measurements presented here as the `Lupus' sample, the \citet{tripathi17} collection as the `SMA' sample, and their combination as the `joint' sample.  Noting the overlap of one target (IM Lup) between the Lupus and SMA samples, the joint sample has a total count of 105 unique disk targets.

\section{Analysis \label{sec:analysis}}

\subsection{Modeling Surface Brightness Profiles \label{subsec:modeling}}

To measure luminosities and sizes for the continuum emission, we follow the procedure described in detail by \citet{tripathi17}.  This approach models the observed interferometric visibilities with a simple parametric prescription for the radial surface brightness profile, and then derives a constraint on an ``effective" size, defined as the radius that encircles a fixed fraction of the luminosity.  \citet{tripathi17} discussed how a size defined in this way is robust against the intrinsic parameter uncertainties in the adopted model, and also the actual choice of model.  Any prescription that describes the data well suggests the same effective size. 

We adopt a Gaussian likelihood that compares the observed complex visibilities and Hankel transforms of ``Nuker" surface brightness profiles \citep{lauer95},
\begin{equation}
    I_{\nu}(\varrho) \propto \left( \frac{\varrho}{\varrho_t} \right)^{-\gamma} \left[ 1 + \left (\frac{\varrho}{\varrho_t} \right) ^{\alpha} \right ] ^{(\gamma - \beta)/\alpha},
    \label{eq:nuker}
\end{equation}
that are shifted for positional offsets from the phase center ($dx$, $dy$), stretched to account for an inclined viewing angle ($i$), and rotated by a sky-plane position angle ($\varphi$).  Equation~\ref{eq:nuker} is normalized based on a total flux density parameter ($F_\nu$, equivalent to its integral over all angular separations $\varrho$ and polar angles).  The model prescription has nine parameters, five in the Nuker profile and four from the geometry of the sky projection.  Adopting the same practical and conservative priors advocated by \citet{tripathi17}, we sample the posterior distribution of the model parameters conditioned on the data using {\tt emcee} \citep{foreman-mackey13}, a Monte Carlo Markov Chain (MCMC) code that employs the affine-invariant ensemble sampler developed by \citet{goodman10}.  A detailed practical description and demonstration of the mechanics of this modeling are provided by \citet{tripathi17}; they are broadly applicable to the analysis performed for this sample.  

Posterior samples of the effective size, $\varrho_{\rm eff}$, are then calculated from the cumulative intensity profile,
\begin{equation}
f_{\nu}(\varrho) = 2 \pi \int_0^{\varrho} I_{\nu}(\varrho^\prime) \, \varrho^\prime \, d\varrho^\prime,
\label{eq:fcum}
\end{equation}
such that $f_\nu(\varrho_{\rm eff}) = x F_\nu$; we adopt $x = 0.68$ (see the discussion by \citealt{tripathi17} and Section~\ref{subsec:sizelum}).  Then, we compute posterior samples for related quantities in physical units, including the effective size, $R_{\rm eff} = \varrho_{\rm eff} \times d$, and the continuum luminosity\footnote{The continuum luminosities are cast here in units of Jy for a distance of 140\,pc, to ease comparisons with other datasets.}, $L_{\rm mm} = F_\nu \times (d/140)^2 \times s$, where $d$ is a distance in pc and $s$ is a multiplicative factor that crudely accounts for a systematic flux calibration uncertainty ($s$ values are drawn from a normal distribution with a mean of unity and a standard deviation of 0.1).  Posterior draws for the distance were made following \citet{astraatmadja16}, conditioned on parallax measurements and assuming an uniform $d$ prior.  Trigonometric parallax ($\varpi$) measurements were collected from the {\it Gaia} DR2 catalog \citep{gaia_dr2} for all but 4 targets in the Lupus sample.\footnote{2MASS J15450634-3417378, 2MASS J16011549-4152351, 2MASS J16070384-3911113, and 2MASS J16075475-3915446.}  In those remaining cases, we assigned $\varpi$ based on the weighted mean of the $\sim$20 nearest Lupus members within a 30\arcmin\ distance; the standard deviation of those neighbor parallaxes was adopted as the associated uncertainty.  For all targets, we corrected the catalog parallaxes for the known 30\,$\mu$as systematic shift with respect to the quasar reference frame, and folded in a 0.1\,mas systematic uncertainty in quadrature with the formal quoted uncertainty \citep[cf.,][]{lindegren18}. 

For the SMA sample, we adopted the parameters inferred by \citet{tripathi17}\footnote{We used the modified results for the UZ Tau E disk computed by \citet{tripathi18}.} and then re-calculated the posteriors on \{$L_{\rm mm}$, $R_{\rm eff}$\} using parallaxes from the {\it Gaia} DR2 catalog \citep{gaia_dr2}.

\subsection{Modifications for Special Cases \label{subsec:special}}

There are 10 targets\footnote{2MASS J16000060-4221567, Sz 130, 2MASS J16070384-3911113, 2MASS J16073773-3921388, Sz 95, Sz 104, 2MASS J16084940-3905393, 2MASS J16085373-3914367, 2MASS J16101984-3836065, and 2MASS J16134410-3736462.} in the Lupus sample for which the MCMC sampling would not converge with the standard priors.  Each of these targets exhibits a faint ($\lesssim$5\,mJy), unresolved emission distribution.  Their relatively noisy visibility data means that we cannot discriminate models well along the degeneracy between the transition radius, $\varrho_t$, and the inner disk index, $\gamma$.  The marginal posteriors on $\varrho_t$ are clearly peaked at low values, but with non-negligible tails that spanned the full prior range (out to 10\arcsec) when $\gamma$ values were pegged to the high end of its prior distribution ($\gamma \approx 2$).  This is a classic problem in modeling disk emission with limited resolution and sensitivity: large sizes can be accommodated by steep gradients  \citep[e.g.,][]{mundy96,aw07a}.  While we cannot rule out the possibility that these faint disks do have systematically steeper brightness profiles, the brighter disks in this sample (and the SMA sample; see \citealt{tripathi17}) where $\gamma$ can be measured reliably favor lower values.  Therefore, we prefer the low--$\gamma$, low--$\varrho_t$ solution to this problem.  To mitigate this issue, we modified the $\varrho_t$ prior for these systems, employing a Gaussian prior with a mean of zero and a width ($\sigma = 0\farcs13$) corresponding to the (naturally weighted) data resolution.  

We also found the need to significantly relax the priors on the index parameters $\beta$ and $\gamma$ for the specific case of the Sz 91 disk, which appears as an extremely narrow belt of emission \citep[for details, see][]{tsukagoshi17}.  We increased the upper bound of the $\beta$ prior from 10 to 20, and extended the low-end turnover of the logistic function prior on $\gamma$ from -3 to -20. 

Two wide binary systems, Sz~65+Sz~66 and 2MASS J16085324-3914401+2MASS~J16085373-3914367, have their individual components observed in distinct ALMA pointings.  However, both components lie within the primary beam for each of those pointings.  In these cases, we only modeled the visibility data from the one pointing centered on each individual component.  A combined modeling of both pointings of each system together would require a reliable treatment of primary beam attenuation, and thereby makes the modeling unnecessarily complicated.  Each pointing has more than sufficient signal that the $\sim$20--30\%\ decrease in the noise level was deemed inconsequential.  However, the companion disk emission in the periphery of each field could bias the modeling.  So, that emission was removed before modeling by subtracting off the Fourier transform of its associated {\tt CLEAN} components.

\begin{figure}[!t]
\includegraphics[width=\linewidth]{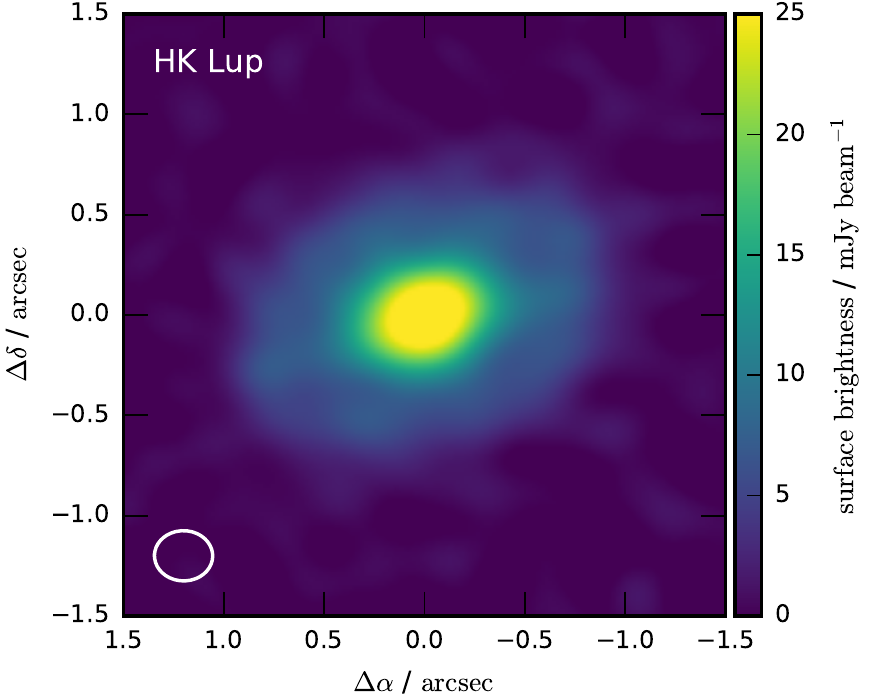}
\figcaption{Synthesized image of the 335\,GHz emission from the HK Lup disk (beam size of $0\farcs29 \times 0\farcs25$, PA = 89\degr) showing a narrow, apparently shallow, depression near a (projected) radius of 0\farcs4 from the peak.  That depression motivates a slight modification of the standard model for this source.   \label{fig:hklup}}
\end{figure}

Finally, we had to modify the model parameterization itself for the HK Lup (Sz 98) disk.  An initial modeling as described above cannot provide a good match to the observed visibilities, which feature a pronounced oscillation indicative of abrupt brightness changes on relatively small angular scales.  A close examination of a continuum image made to enhance the resolution (with Briggs robust=0), as shown in Figure~\ref{fig:hklup}, indicates a shallow depression in the intensity distribution at a (deprojected) radius of $\sim$0\farcs4.  With the image as a guide, we modified the Nuker brightness profile model to include three additional parameters that can mimic this depression: an inner radius ($\varrho_d$), a width ($\Delta \varrho_d$), and a depletion factor ($\epsilon_d$), such that the functional form in Equation~\ref{eq:nuker} is multiplied by $\epsilon_d$ when $\varrho \in (\varrho_d$, $\varrho_d + \Delta \varrho_d)$.  Because the depression is at best marginally resolved, we set somewhat restrictive Gaussian priors on the spatial parameters, $p(\varrho_d) \sim \mathcal{N}(0.35, 0.10)$ and $p(\Delta \varrho_d) \sim \mathcal{N}(0.35, 0.10)$.  The prior on the depletion factor is less stringent, with $p(\log{\epsilon_d}) \sim \mathcal{U}(-4, 0)$.  Unsurprisingly, this modified profile provides a much improved fit to the visibility data, and thereby a more reliable $\varrho_{\rm eff}$.  We find $\varrho_d = 0\farcs36\pm0.03$, $\Delta \varrho_d = 0\farcs43\pm0.05$, and $\log{\epsilon_d} = -0.4\pm0.1$, where the quoted values are the peaks of the marginal posteriors and the 68.3\%\ confidence interval ranges (the other parameter values will be presented in Section~\ref{subsec:results} with the rest of the sample).  These should be considered tentative values, since the data resolution is not really sufficient to do a proper evaluation of the depression properties (i.e., there is a propensity to over-fit in this case).  In the end, because the perturbation is modest, we end up with the same effective size ($\varrho_{\rm eff}$) as the depression-free model.  

\subsection{Stellar and Accretion Parameter Estimation \label{subsec:stars}}

A variety of demographics inquiries that follow require estimates of some basic stellar and accretion parameters.  Nearly all of the Lupus sample host stars have been studied with extensive spectroscopic analyses in a dedicated campaign with the X-shooter spectrograph at the VLT \citep{alcala14,alcala17,frasca17,biazzo17}.  We primarily use the parameters compiled by \citet{alcala17} as a base set of values, but do make some modifications as appropriate.  In their Table A.2, \citeauthor{alcala17}~list effective temperatures ($T_{\rm eff}$) and stellar luminosities ($L_\ast$), and their associated uncertainties, for a presumed (fixed) distance.  The $L_\ast$ measurements are quoted in the linear domain, but presumably this is for clarity of tabulation and the inferences are actually made in something equivalent to $\log L_\ast$ (since the quoted uncertainties would frequently imply physically impossible negative values).  We re-cast their estimates into their corresponding $\log{T_{\rm eff}}$ and $\log{L_\ast}$ posteriors, where the latter is properly scaled to account for our adopted $d$ posteriors (Section~\ref{subsec:modeling}).\footnote{For $T_{\rm eff}$, we draw random samples from a normal distribution with the specified means and standard deviations and take their logarithms.  For $L_\ast$, we estimate an appropriate standard deviation in $\log{L_\ast}$ based on the specified uncertainty and logarithm of the mean, and then proceed as for $T_{\rm eff}$.  For each $\log{L_star}$ draw, we add an appropriate term to account for the $d$ re-scaling.}

We then compare these measurements to pre-main sequence evolutionary models in the Hertzsprung-Russell (H-R) diagram to make estimates of the target masses and ages.  To do that, we assume that $T_{\rm eff}$ and $L_\ast$ are independent (this should produce a conservative bias that in a sense over-estimates the uncertainties on $M_\ast$ and $\tau_\ast$) and adopt the approach advocated by \citet{jorgensen05}, and implemented in the {\tt ScottiePippen} software package \citep[see][]{czekala16}, to sample the joint posterior distribution of \{$M_\ast$, $\tau_\ast$\} conditioned on \{$T_{\rm eff}$, $L_\ast$\}.  We prefer here the MIST evolutionary models \citep{choi16} as the basis for that inference, since they are well-sampled across the parameter-space relevant for these targets.  The same analysis was performed for various other evolutionary models \citep{siess00,dotter08,tognelli11,baraffe15} and no significant differences were identified.  

Accretion rates were calculated following \citet{alcala17}, based on their estimates of the accretion luminosities, $L_{\rm acc}$ (their Table A.3).  When a measurement was available, their tabulated values were used to construct log-normal $L_{\rm acc}$ distributions with their suggested dispersion of 0.25\,dex applied uniformly.  When only upper limits are quoted, we presumed that $\log{L_{\rm acc}}$ is distributed like $\sim{\rm erfc}(\log{L_{\rm acc}} - {\rm UL})/0.25$, where UL is the quoted upper limit on $\log{L_{\rm acc}}$ and erfc is the complementary error function \citep[e.g., see][]{law17}.  In all cases, the input $L_{\rm acc}$ was scaled to account for the $d$ distributions adopted here.  The derived $\log{\dot{M_\ast}}$ posteriors were then calculated by combining those distributions with the inferences on $M_\ast$, $L_\ast$, and $T_{\rm eff}$ \citep[where the latter two terms define the stellar radius; see][their Equation~1, for details]{alcala17}.  It is worth noting that the cases which only produce upper (lower) limits on $M_\ast$ will correspond to lower (upper) limits on $\dot{M_\ast}$.    

The same analysis was performed for the SMA sample targets, with measurements collated from various literature sources (references are provided below).

\section{Results \label{sec:results}}

\subsection{Surface Brightness Modeling Results \label{subsec:results}}

In an effort to keep the focus on the demographics, we have collated the detailed individual results in Appendix~\ref{appendixA}.  The inferred model parameters for the Lupus sample disks are summarized in Table~\ref{table:SBpars}, which lists the peaks of the marginal posteriors and the 68.3\%\ confidence interval ranges (or the 95.5\%\ confidence limits, as appropriate).  The equivalent summaries for the SMA sample are provided in Table~\ref{table:SBpars2}.  These tables also include posterior summaries for the derived effective sizes, $\varrho_{\rm eff}$, as well as for $\log{L_{\rm mm}}$ and $\log{R_{\rm eff}}$ (see Section~\ref{subsec:modeling}).  For the disks in the Lupus sample, Figure~\ref{fig:visgallery} compares the observed visibilities with synthetic data generated from random draws from the parameter posterior distributions.  Figures~\ref{fig:profilesA} and \ref{fig:profilesB} show the inferred surface brightness profiles ($I_\nu$; Equation~\ref{eq:nuker}) and cumulative intensity profiles ($f_\nu$; Equation~\ref{eq:fcum}).  The corresponding results for the SMA sample are shown by \citet{tripathi17} in their Figures~7 and 8, respectively.

\begin{figure}[t!]
\includegraphics[width=\linewidth]{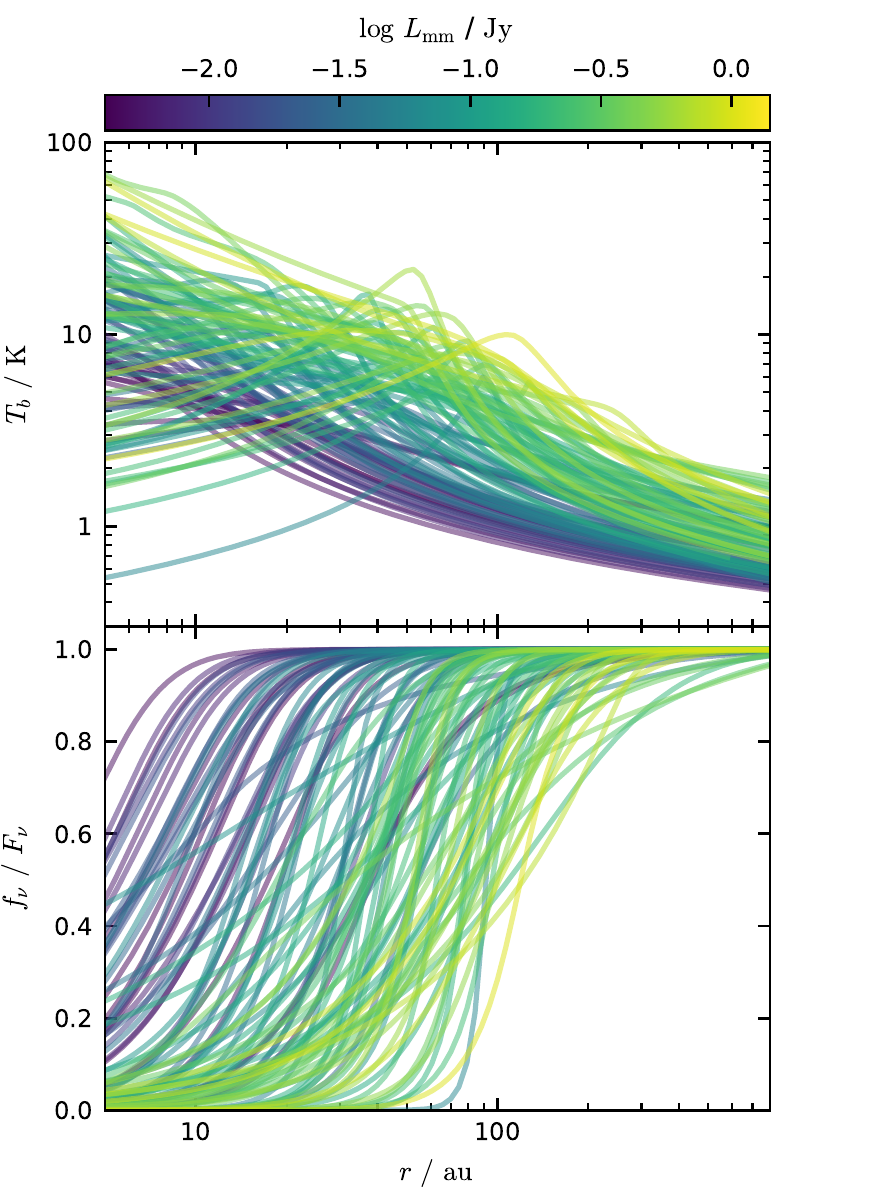}
\figcaption{Brightness temperature ({\it top}) and cumulative intensity ({\it bottom}) radial profiles for each disk in the joint sample, in each case defined as the median profile constructed from samples of the parameter posterior distributions.  The color-coding tracks $L_{\rm mm}$; a scale-bar is shown at the top of the figure.     
\label{fig:all_profiles}}
\end{figure}

Figure~\ref{fig:all_profiles} effectively summarizes the brightness profile modeling for all 105 disks in the joint sample.  It shows the brightness temperature ($T_b$; computed from the full Planck function) and (normalized) cumulative intensity profiles ($f_\nu/F_\nu$), constructed from the medians of posterior surface brightness values at each radius.  Each profile is color-coded by its $\log{L_{\rm mm}}$ value (see Tables~\ref{table:SBpars} and \ref{table:SBpars2}).  Before exploring these results in more detail below, there are a few basic features worth mentioning from these plots.  First, the $T_b$ profiles actually show relatively small dispersion.  The median $T_b$ at 40\,au is $\sim$5\,K, with a standard deviation of only a factor of two despite the more than two orders of magnitude spread in $L_{\rm mm}$.  Second, the $T_b$ values are lower than would be expected for typical dust temperatures, but not considerably so (certainly within a factor of 5--10).  This suggests either that the continuum is marginally optically thin ($\tau_\nu \gtrsim 0.1$), or that it could be optically thick if the true emission distribution is poorly resolved (i.e., there is structure on scales $\gtrsim$2$\times$ smaller than the resolution).  And third, the color-coding of the $f_\nu/F_\nu$ profiles clearly demonstrates the presence of a size--luminosity relationship: more luminous disks have more emission distributed at larger distances from their host stars.           

Based on the modeling results (and the appearance of the data), we find that 9 of the 56 Lupus targets exhibit a large central depression in their emission profiles: Sz 84, RY Lup, 2MASS J16070854-3914075, Sz 91, Sz 100, 2MASS J16083070-3828268, Sz 111, 2MASS J16090141-3925119, and Sz 118.  Most of these were already identified and studied in more detail by \citet{vandermarel18}.  With slightly different selection criteria, their sample does not include 2MASS J16090141-3925119 (although we note that \citealt{ansdell16} tentatively identified a central depression in its associated emission), but adds in Sz 123 A, 2MASS J16102955-3922144, and MY Lup.  We concur that Sz 123 A has a central depression, but we exclude it here due to our criterion against selecting targets with close companions.  We do not find any evidence that 2MASS J16102955-3922144 or MY Lup have centrally depressed emission morphologies.  There are an additional 17 such targets in the SMA sample: LkH$\alpha$ 330, IRAS 04125+2902, UX Tau A, DM Tau, LkCa 15, GM Aur, MWC 758, CQ Tau, TW Hya, SAO 206462, RX J1604.3-2130, RX J1615.3-3255, SR 24 S, SR 21, WSB 60, DoAr 44, and RX J1633.9-2442.  In the following, we will refer to these targets as ``transition disks".  Despite the \citet{tripathi17} selection bias (caused by folding in the targeted SMA transition disk survey by \citealt{andrews11}), the total fraction of transition disks in the joint sample, $\sim$25\%, is similar to what is found in infrared surveys of other $\sim$Myr-old clusters \citep[e.g., see][]{currie11}.

\subsection{Host Parameters \label{subsec:hosts}}

\begin{figure}[t!]
\includegraphics[width=\linewidth]{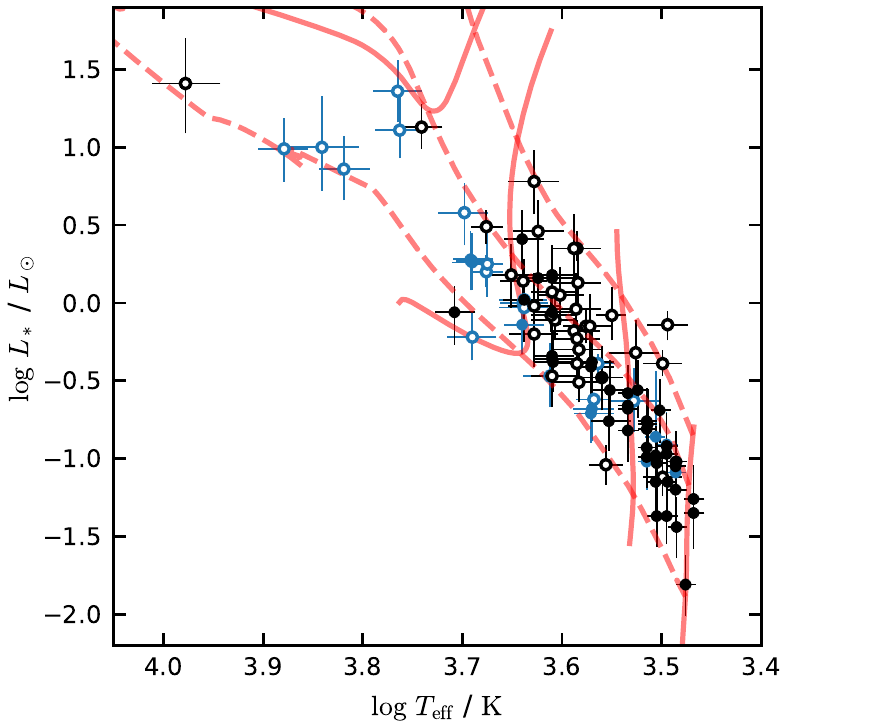}
\figcaption{The H-R diagram for the joint sample.  The Lupus and SMA samples are shown with closed and open markers, respectively; transition disk hosts are colored blue.  The solid curves show MIST evolutionary tracks for ({\it from top left to bottom right}) $M_\ast/M_\odot$ = 3, 1, 0.3, 0.1; the dashed curves show ({\it from right to left}) the $\tau_\ast$/Myr = 0.1, 1, 10 isochrones \citep{choi16}.  
\label{fig:hrd}} 
\end{figure}

The inferred stellar host properties, including the distances used to calculate some of the continuum emission properties in physical units, are summarized in Tables~\ref{table:stars} and \ref{table:starsB} for the Lupus and SMA samples, respectively.  Figure~\ref{fig:hrd} shows the location of the target hosts in the H-R diagram, with some representative MIST evolutionary models and isochrones overlaid \citep{choi16}.  The joint sample spans the full mass range of T Tauri and Herbig AeBe stars, from 0.1 to 3\,$M_\odot$, although the vast majority has $M_\ast \lesssim 1$\,$M_\odot$.  The mean age of the sample is about 1.5\,Myr, although with a wide dispersion (the combined posterior for all sample targets has a peak at $\log{\tau_\ast/{\rm yr}} \approx 6.2$, with 68.3\%\ of the posterior probability within $\pm$0.5\,dex of that value).  There are no discernible differences in the age distributions of the Lupus and SMA samples.  \dt{The latter has a $M_\ast$ distribution relatively biased toward the high end, but on the mass range where both samples overlap ($-0.6 \le \log{M_\ast} \le 0.2$) their relative mass functions are indistinguishable.}

\dt{In much of the following analysis, we will consider the joint sample under the implicit assumption that the parent distributions of the host properties for sources in Lupus and the clusters from which the SMA sample are drawn (primarily Taurus and Ophiuchus) are the same.  Ultimately that is an assumption that ought to be revisited when complete samples of spatially resolved continuum measurements are available for those clusters.}

\begin{deluxetable*}{cccccc|c}[t!]
\tablecaption{Selected Regression Posterior Summaries\label{table:regressions}}
\tablehead{
\colhead{$\mathsf{x}$} &
\colhead{$\mathsf{y}$} &
\colhead{sample} & 
\colhead{$\mathcal{A}$} &
\colhead{$\mathcal{B}$} &
\colhead{$\sigma$} & 
\colhead{$\hat{\rho}$, CI$_\rho$(99\%)}
}
\startdata
$\log{L_{\rm mm}/{\rm Jy}}$ & $\log{R_{\rm eff}/{\rm au}}$ & Lupus 
     & \phd2.15 $^{+0.10}_{-0.10}$ & 0.51 $^{+0.07}_{-0.06}$ & 0.25 $^{+0.04}_{-0.03}$ 
     & 0.81 (0.60, 0.94) \\
$\log{L_{\rm mm}/{\rm Jy}}$ & $\log{R_{\rm eff}/{\rm au}}$ & SMA 
     & \phd2.10 $^{+0.05}_{-0.05}$ & 0.48 $^{+0.06}_{-0.08}$ & 0.17 $^{+0.03}_{-0.01}$ 
     & 0.77 (0.53, 0.97) \\
$\log{L_{\rm mm}/{\rm Jy}}$ & $\log{R_{\rm eff}/{\rm au}}$ & joint 
     & \phd2.10 $^{+0.06}_{-0.03}$ & 0.49 $^{+0.05}_{-0.03}$ & 0.20 $^{+0.02}_{-0.01}$ 
     & 0.86 (0.76, 0.92) \\
\hline
$\log{L_\ast/L_\odot}$ & $\log{L_{\rm mm}/{\rm Jy}}$ & joint
     & -0.88 $^{+0.06}_{-0.06}$ & 0.84 $^{+0.08}_{-0.08}$ & 0.47 $^{+0.04}_{-0.03}$ 
     & 0.75 (0.59, 0.90) \\
$\log{M_\ast/M_\odot}$ & $\log{L_{\rm mm}/{\rm Jy}}$ & joint
     & -0.69 $^{+0.08}_{-0.05}$ & 1.51 $^{+0.11}_{-0.16}$ & 0.46 $^{+0.04}_{-0.03}$ 
     & 0.75 (0.61, 0.86) \\
$\log{L_\ast/L_\odot}$ & $\log{R_{\rm eff}/{\rm au}}$ & joint
     & \phd1.69 $^{+0.04}_{-0.04}$ & 0.30 $^{+0.06}_{-0.05}$ & 0.31 $^{+0.03}_{-0.03}$ 
     & 0.54 (0.29, 0.81) \\
$\log{M_\ast/M_\odot}$ & $\log{R_{\rm eff}/{\rm au}}$ & joint
     & \phd1.77 $^{+0.05}_{-0.04}$ & 0.58 $^{+0.10}_{-0.10}$ & 0.30 $^{+0.03}_{-0.03}$ 
     & 0.58 (0.35, 0.74) \\
\hline
$\log{L_{\rm mm}/{\rm Jy}}$ & $\log{\dot{M}_\ast/M_\odot {\rm yr}^{-1}}$ & joint 
     & \phn-7.6 $^{+0.2}_{-0.2}$ & 1.03 $^{+0.15}_{-0.13}$ & 0.80 $^{+0.10}_{-0.07}$ 
     & 0.67 (0.47, 0.82) \\
$\log{R_{\rm eff}/{\rm au}}$ & $\log{\dot{M}_\ast/M_\odot {\rm yr}^{-1}}$ & joint
     & -10.9 $^{+0.7}_{-0.5}$ & 1.25 $^{+0.35}_{-0.41}$ & 1.04 $^{+0.12}_{-0.08}$ 
     & 0.36 (0.06, 0.64) \\
$\log{L_\ast/L_\odot}$ & $\log{\dot{M}_\ast/M_\odot {\rm yr}^{-1}}$ & joint 
     & \phn-8.3 $^{+0.1}_{-0.1}$ & 1.42 $^{+0.14}_{-0.15}$ & 0.66 $^{+0.08}_{-0.08}$ 
     & 0.81 (0.64, 0.94) \\
$\log{M_\ast/M_\odot}$ & $\log{\dot{M}_\ast/M_\odot {\rm yr}^{-1}}$ & joint 
     & \phn-8.2 $^{+0.1}_{-0.1}$ & 1.95 $^{+0.25}_{-0.32}$ & 0.84 $^{+0.11}_{-0.07}$ 
     & 0.63 (0.40, 0.79) \\
\enddata
\tablecomments{The quoted values for $\mathcal{A}$, $\mathcal{B}$, and $\sigma$ are the peaks of their posterior distributions; the corresponding uncertainties represent the 68\%\ confidence interval.  The quoted correlation coefficients ($\rho$) and associated confidence intervals (CI$_\rho$) represent the median and 99\%\ confidence regions of the $10^4$ posterior samples for the regression.}
\end{deluxetable*}

\subsection{Inferred Scaling Relations \label{subsec:scalings}}

In this section, we aim to quantify the properties of any scaling relations between the mm continuum sizes and luminosities (Section~\ref{subsec:sizelum}), and between those parameters and the host star properties (Section~\ref{subsec:hosts}) and accretion rates (Section~\ref{subsec:accretion}).  For the sake of simplicity, we will interpret any such scaling relations with a linear regression analysis -- thereby presuming power-law behavior when the variables of interest are logarithmic -- following the general mixture model formulation described in detail by \citet{kelly07}.\footnote{We use the same algorithms, but in software modified for {\tt python} by J.~Meyers (\url{https://github.com/jmeyers314/linmix}).}  To succinctly summarize, that regression analysis presumes a relation
\begin{equation}
    \mathsf{y} = \mathcal{A} + \mathcal{B} \, \mathsf{x} + \varepsilon,
\end{equation}
where \{$\mathsf{x}$, $\mathsf{y}$\} are the variables of interest (with associated covariances), $\mathcal{A}$ and $\mathcal{B}$ are the intercept (normalization) and slope (power-law index) of the model scaling relation, and $\varepsilon$ represents an additional Gaussian ``scatter" (in $\mathsf{y}$) around the mean relation (i.e., $\varepsilon$ is drawn from a Gaussian distribution with mean zero and standard deviation $\sigma$).  The \citet{kelly07} methodology infers posterior samples on \{$\mathcal{A}$, $\mathcal{B}$, $\sigma$\} conditioned on \{$\mathsf{x}$, $\mathsf{y}$\}, their covariance matrices, and any associated censoring (limits).  We have further simplified this analysis by approximating the two-dimensional posteriors for any \{$\mathsf{x}$, $\mathsf{y}$\} as (symmetric, elliptical) Gaussian distributions with representative means and covariances.

\subsubsection{Size--Luminosity Relation: \{$L_{\rm mm}$, $R_{\rm eff}$\} \label{subsec:sizelum}}

\begin{figure}[t!]
\includegraphics[width=\linewidth]{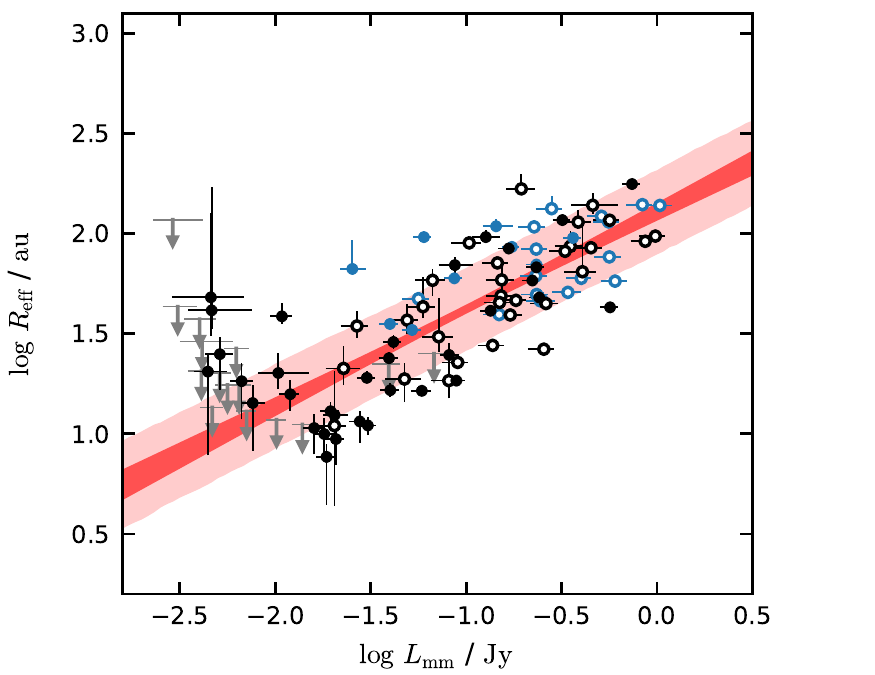}
\figcaption{The mm continuum size--luminosity relationship.  Symbols are as in Figure~\ref{fig:hrd}.  Error bars mark the 68\%\ confidence intervals; upper limits (at 95\%\ confidence) are marked with gray arrows.  The dark shaded region marks the 68\%\ confidence interval on the scaling relation from the linear regression analysis on the joint sample.  The lighter shaded region marks the same behavior with the additional scatter term ($\sigma$) folded in.  
\label{fig:lumsize}}
\end{figure}

\begin{figure*}[t!]
\includegraphics[width=0.5\linewidth]{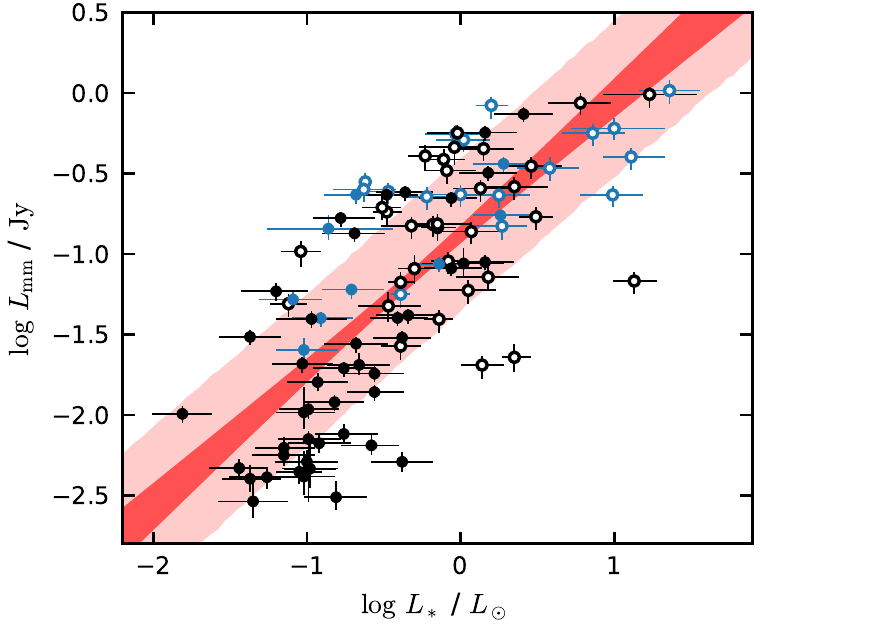}
\includegraphics[width=0.5\linewidth]{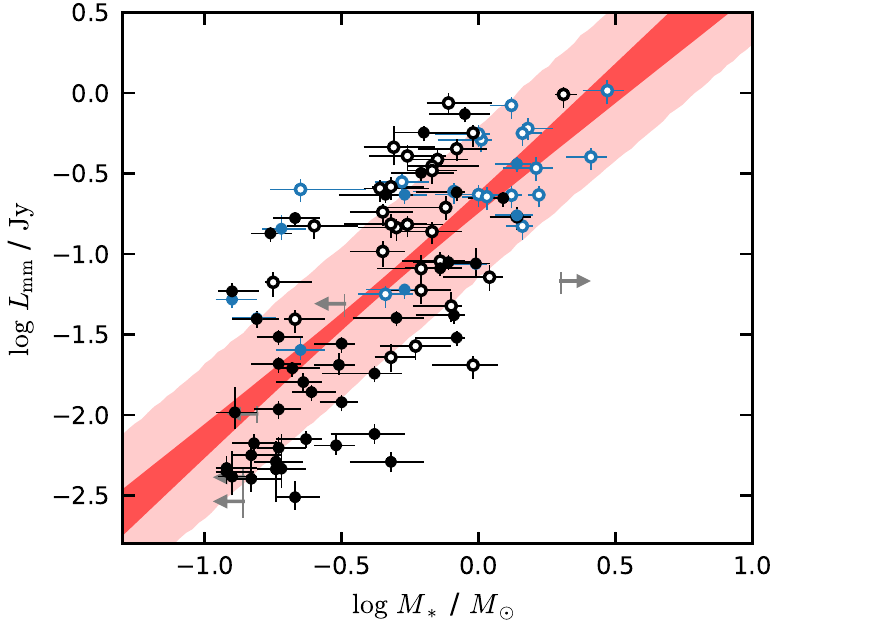}
\includegraphics[width=0.5\linewidth]{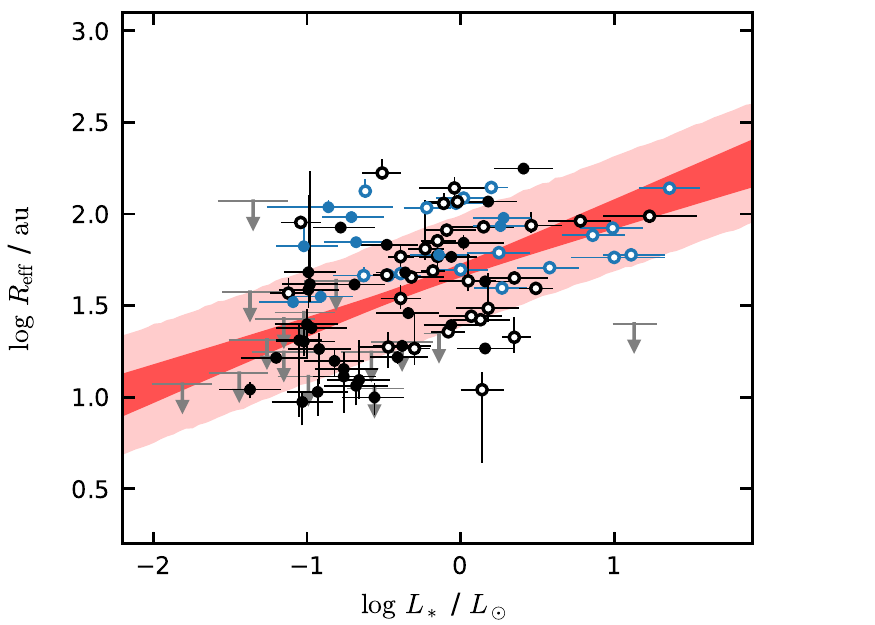}
\includegraphics[width=0.5\linewidth]{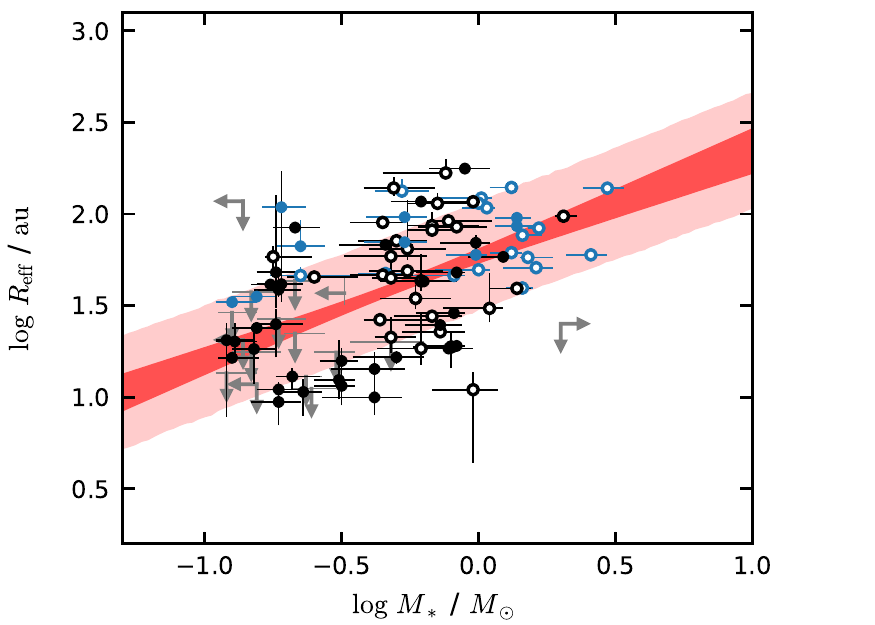}
\figcaption{The top panels show how the continuum luminosity ($L_{\rm mm}$) scales with the host star luminosity ($L_\ast$; left) and mass ($M_\ast$; right).  The bottom panels show the corresponding scalings with the continuum effective size ($R_{\rm eff}$).  Annotations are as in Figure~\ref{fig:lumsize}.  
\label{fig:lmmreff_hosts}}
\end{figure*}

Figure~\ref{fig:lumsize} shows the continuum size--luminosity scaling for the Lupus and SMA samples, along with the corresponding regression results.  A dark shading marks the 68\%\ confidence interval region inferred for the joint sample; the lighter shading includes the scatter term.  We find a definitive size--luminosity relationship in both the joint sample and its constituent sub-samples.  The regression parameters are listed in Table~\ref{table:regressions}.  The inferred size--luminosity scaling is the same within the uncertainties for either sub-sample or their union.  Moreover, we find that the slope of this relationship is identical regardless of how $R_{\rm eff}$ is defined.  The same behavior is inferred if we consider the radii that encircle 50, 68 (our adopted value), 80, 90, or 95\%\ of $L_{\rm mm}$; the normalization (intercept) changes modestly, from $\mathcal{A} \approx 2.0$ for the 50\%\ definition to 2.4 for the 95\%\ definition. 

These results verify the original \citet{tripathi17} conclusions from the SMA sample alone, that there is a clear and linear relationship between the continuum luminosity and its emitting {\it surface area}, $L_{\rm mm} \propto R_{\rm eff}^2$.  The Lupus sample extends that relation almost an order of magnitude lower in $L_{\rm mm}$ (albeit with considerable uncertainty on size estimates there), suggesting that the original analysis from the SMA sample was not substantially biased by either a lack of depth or by combining targets from different cluster environments.  It is notable that the inferred scatter around the mean relation does not appreciably shrink for the joint sample of 105 disks; the implication is that the relation has some intrinsic dispersion and/or higher dimensionality.

\begin{figure*}[t!]
\includegraphics[width=0.5\linewidth]{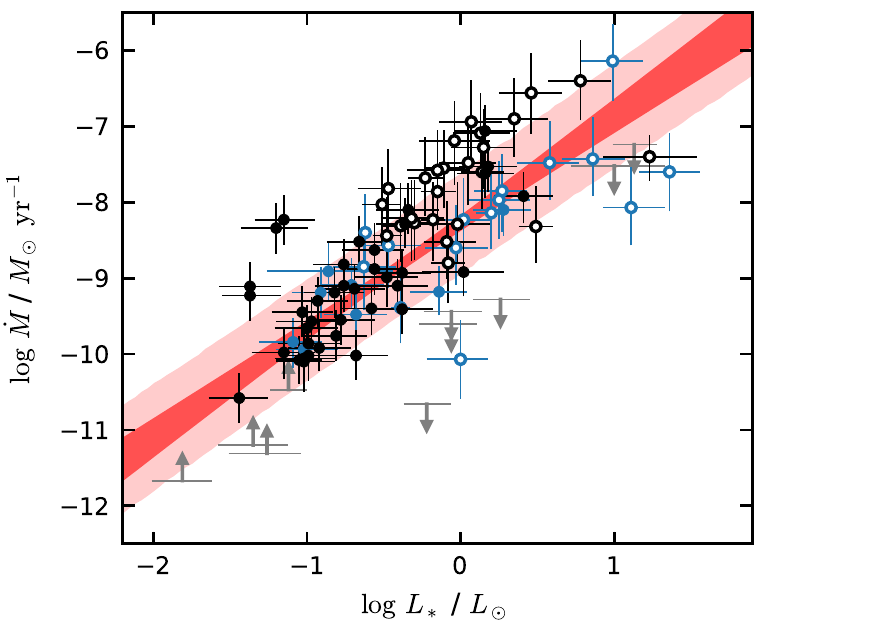}
\includegraphics[width=0.5\linewidth]{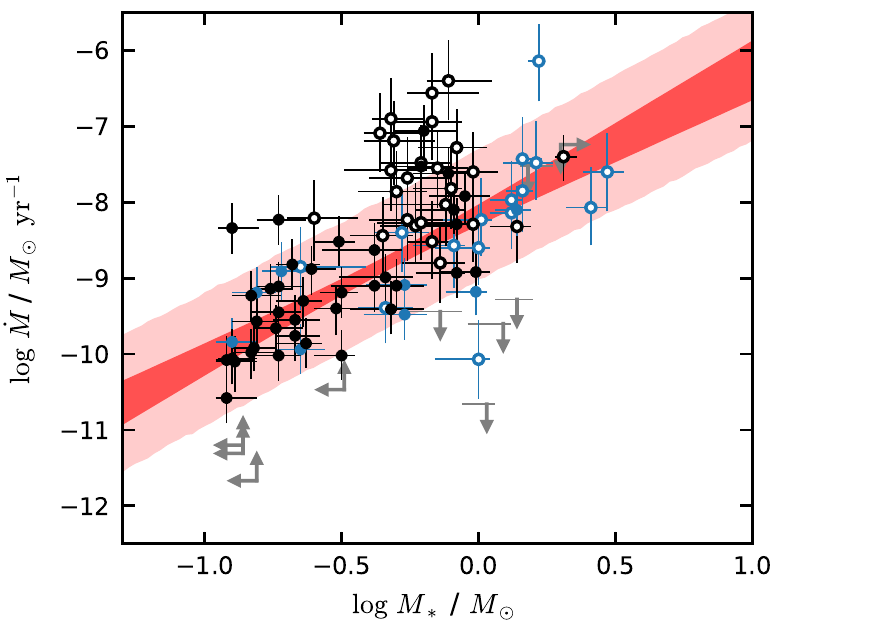}
\includegraphics[width=0.5\linewidth]{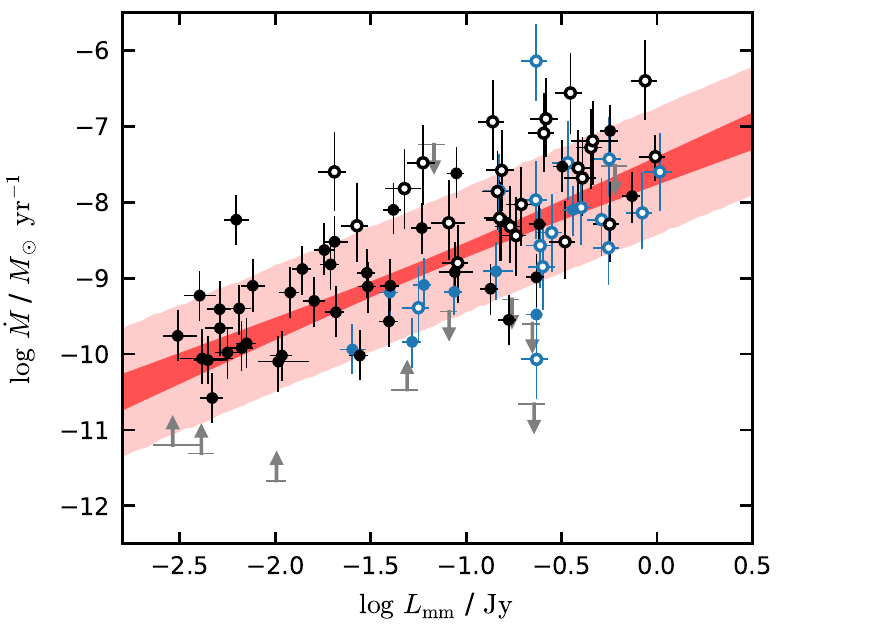}
\includegraphics[width=0.5\linewidth]{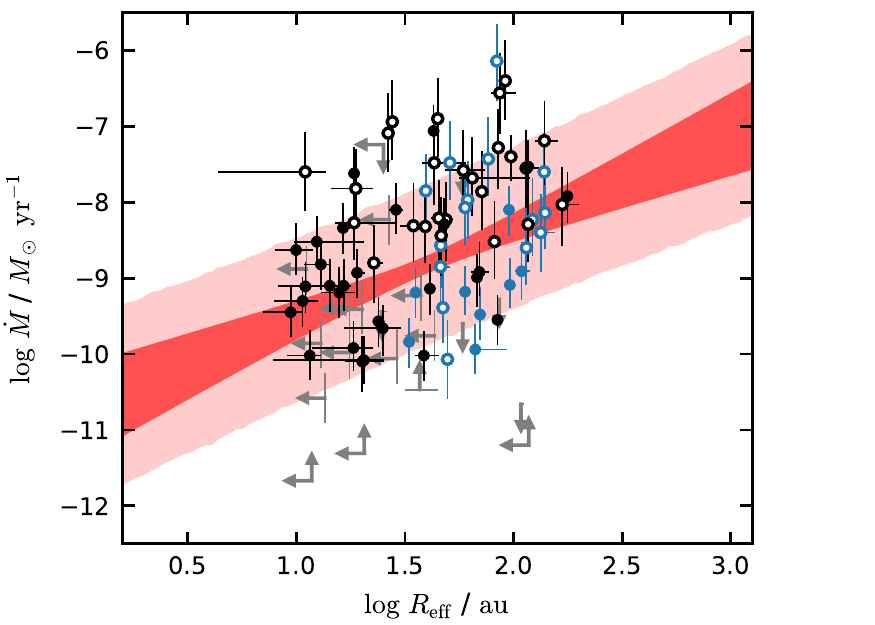}
\figcaption{The top panels show how the accretion rate ($\dot{M}_\ast$) scales with the host star luminosity ($L_\ast$; left) and mass ($M_\ast$; right).  The bottom panels show its behavior as a function of continuum luminosity ($L_{\rm mm}$; left) and size ($R_{\rm eff}$; right).  Annotations are as in Figure~\ref{fig:lumsize}.  
\label{fig:lmmreff_acc}}
\end{figure*}

\dt{As discussed by \citet{tripathi17} (see also Section~\ref{subsec:thick}), disks cannot populate the lower right corner of the $R_{\rm eff}$--$L_{\rm mm}$ plane because emission levels saturate at sizes a factor of $\sim$3 below the mean relation when the continuum becomes optically thick.  There is no obvious physical reason that the region above the mean relation should be depopulated, but it is worth considering two potential selection effects.  First is the possibility that we have excluded large, low-surface brightness disks in the Lupus sample because of the sensitivity criterion in Section~\ref{sec:sample} (i.e., the continuum must be firmly detected).  This can be ruled out.  For any reasonable brightness profile (Nuker profiles with a range of gradient parameters), the \citet{ansdell16} ALMA survey is sensitive enough that we would have included any target with $R_{\rm eff}$ values (at least) 10$\times$ larger than the mean size--luminosity relation.  A second, more subtle, possibility is that the disks that populate the upper left of the size--luminosity plane were excluded from the \citeauthor{ansdell16}~sample because they have negligible (optically thin) infrared excess emission.  Again, for a range of brightness profiles that produce such large $R_{\rm eff}$ for their $L_{\rm mm}$, and reasonable assumptions about dust temperatures and opacities, we find it difficult to accommodate the possibility of optically thin infrared emission.  One interesting exception is a scenario with a low-mass ($\sim$0.1\,$M_\odot$) host and a large ($\gtrsim 70$\,au in radius) dust ring.  Deep imaging surveys of Class III sources could perhaps identify such elusive systems.}

\subsubsection{Disk--Host Relations: \{$[L_{\rm mm}$, $R_{\rm eff}]$, $[L_\ast$, $M_\ast]$\}} \label{subsec:diskhosts}

We consider comparisons of $L_{\rm mm}$ and $R_{\rm eff}$ with the four representative host star properties calculated for the sample targets: effective temperature ($T_{\rm eff}$), luminosity ($L_\ast$), mass ($M_\ast$), and age ($\tau_\ast$).  No relationships were found as a function of age, which is not surprising given the general co-evality of the sample and the large individual $\tau_\ast$ uncertainties (i.e., there is simply not enough dynamic range in the age dimension for this sample).  And while there are significant scalings between these disk properties and $T_{\rm eff}$, we find those relationships are more readily interpreted in the contexts of $M_\ast$ (a natural proxy for $T_{\rm eff}$ for a young, roughly co-eval population).  Therefore, we will focus the discussion on connections between \{$L_{\rm mm}$, $R_{\rm eff}$\} and \{$L_\ast$, $M_\ast$\}.      

We find clear correlations of $L_{\rm mm}$ with either $L_\ast$ or $M_\ast$.  The derived regression parameters are summarized in Table~\ref{table:regressions}; the scaling relations are shown in the top panels of Figure~\ref{fig:lmmreff_hosts}.  The regression analyses for the joint sample suggests that $L_{\rm mm} \propto L_\ast^{0.8}$ and $\propto M_\ast^{1.5}$, with similar dispersions ($\sim$0.5\,dex in $L_{\rm mm}$).  Those scalings make sense with respect to one another, given the approximate mass-luminosity relation for the $\sim$Myr isochrones in the evolution models (roughly $L_\ast \propto M_\ast^{1.8}$).  The same analysis of the Lupus sample alone finds similar results.  That is not true of the SMA sample only, where there is not evidence for a correlation due to a selection bias toward more luminous (therefore massive) hosts.  The $L_{\rm mm}-M_\ast$ relation derived here is consistent with those found from previous continuum photometry surveys of comparable size (and some overlapping targets) and age in the Taurus \citep{andrews13}, Lupus \citep{ansdell16}, and Cha~I \citep{pascucci16} populations.  That said, it is not clear in Figure~\ref{fig:lmmreff_hosts} that these scalings are optimally described by single power-laws: the least luminous (massive) hosts tend to have (marginally) lower $L_{\rm mm}$ than the regression trends would suggest.  

The inferred relationships between $R_{\rm eff}$ and $L_\ast$ or $M_\ast$ are less robust (correlation coefficients $\sim$0.5--0.6), but in line with expectations given the continuum size--luminosity scaling and the connections between $L_{\rm mm}$ and host properties.  Those relationships are shown in the bottom panels of Figure~\ref{fig:lmmreff_hosts} and summarized in Table~\ref{table:regressions}: they suggest that $R_{\rm eff} \propto L_\ast^{0.3}$ and $\propto M_\ast^{0.6}$, with a dispersion of $\sim$0.3\,dex in $R_{\rm eff}$ around those scalings.  These correlations are not found in the SMA sample (again, due to the same selection bias outlined above), but are present in the Lupus sample alone.  We are unaware of any other claims for relationships between the continuum size and $L_\ast$ or $M_\ast$ in the literature.

\subsubsection{Links to Accretion: \{$[L_{\rm mm}$, $R_{\rm eff}]$, $[L_\ast$, $M_\ast]$, $\dot{M}_\ast$\}} \label{subsec:accretion}

Figure~\ref{fig:lmmreff_acc} explores the relationships between the accretion rates and both the stellar host parameters and disk continuum emission parameters for this sample.  The corresponding regression parameters are also included in Table~\ref{table:regressions}.  We find strong correlations between $\dot{M}_\ast$ and the host parameters, such that $\dot{M}_\ast \propto L_\ast^{1.4}$ or $\propto M_\ast^{2.0}$, albeit with considerable scatter (0.7--0.8\,dex in $\dot{M}_\ast$).  This behavior agrees well with many previous studies of the mass-dependence of accretion rates \citep[e.g.,][]{muzerolle03,herczeg08,rigliaco12,alcala14,manara15}.  

We also find a clear {\it linear} correlation between the accretion rate and continuum luminosity, with a dispersion of $\sim$0.8\,dex in $\dot{M}_\ast$.  For a simple scaling between $L_{\rm mm}$ and the disk mass, this behavior is consistent with such relationships found recently by \citet{manara16} and \citet{mulders17}.  There is only a relatively weak suggestion for a roughly linear relationship between accretion rates and continuum sizes, with considerable scatter ($\sim$1\,dex in $\dot{M}_\ast$).  

\begin{figure*}[t!]
\includegraphics[width=\linewidth]{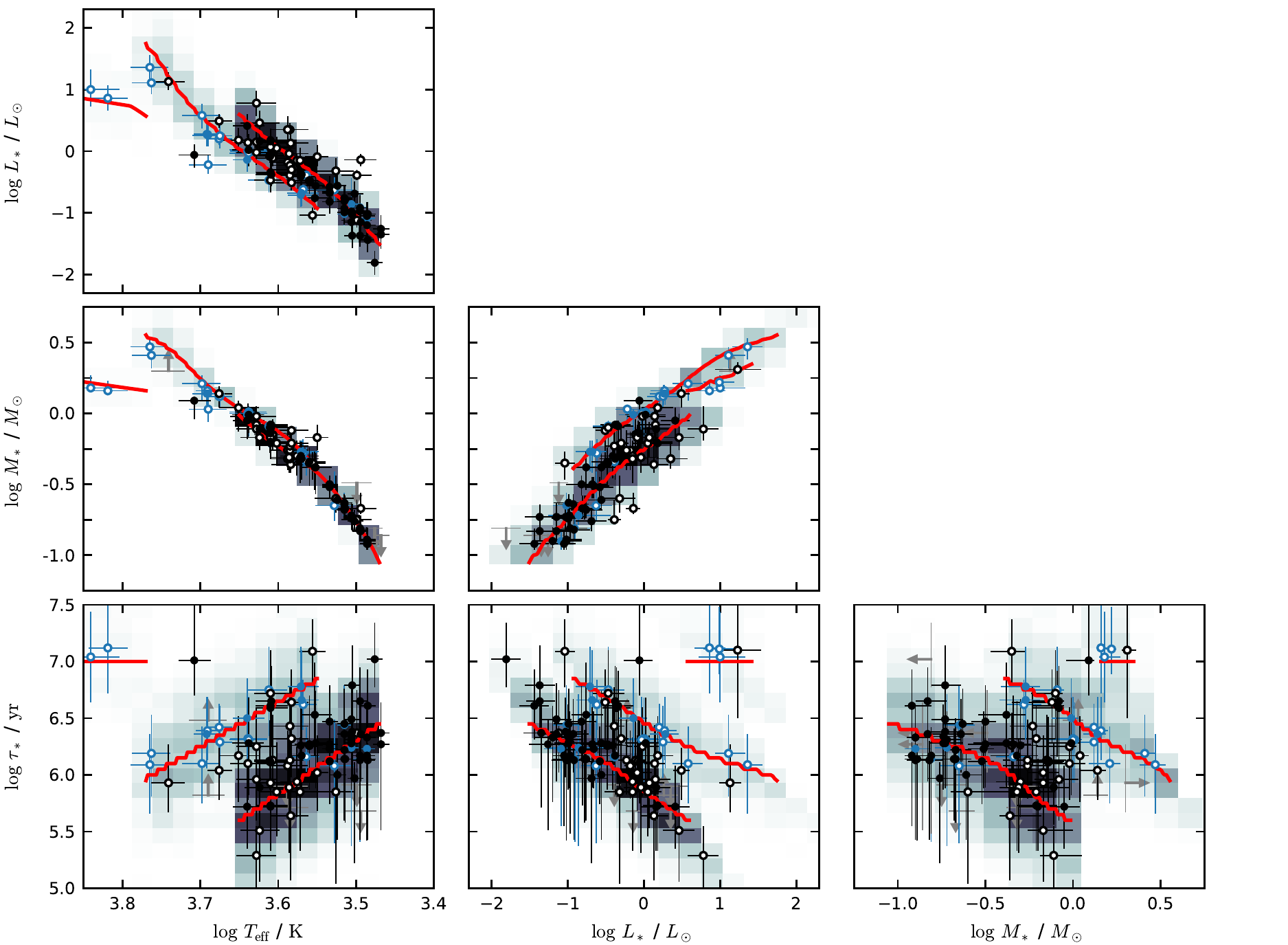}
\figcaption{The relationships between stellar host parameters for the joint sample (points, symbols as in Figure~\ref{fig:hrd}.  The red curves describe the mean behavior of a population model, manually designed to match the sample (see Appendix~\ref{appendixB}).  The grayscale represents a probability density map, folding in an appropriate amount of scatter to account for the measured dispersion in the stellar parameters. 
\label{fig:stars_covar}
}
\end{figure*}

\section{Discussion \label{sec:discussion}}

We have used archival ALMA and SMA observations to measure the sizes ($R_{\rm eff}$) and luminosities ($L_{\rm mm}$) of the 340\,GHz continuum emission from 105 nearby protoplanetary disks.  After collating measurements of the effective temperatures, luminosities, and accretion luminosities of the corresponding stellar hosts from the literature, we computed a homogenized set of host masses, ages, and accretion rates and then examined the multi-dimensional relationships among these basic demographic parameters.  With this analysis, we confirm the $L_\ast$--$L_{\rm mm}$ (or, equivalently, $M_\ast$--$L_{\rm mm}$) scaling relations inferred previously in various $\sim$Myr-old young clusters \citep{andrews13,ansdell16,pascucci16}, the recent claims of a $L_{\rm mm}$--$\dot{M}_\ast$ correlation \citep{manara16,mulders17}, and the classical $M_\ast$--$\dot{M}_\ast$ scaling \citep[e.g.,][]{muzerolle03,manara15}.  Moreover, we verify the $L_{\rm mm}$--$R_{\rm eff}$ scaling inferred by \citet{tripathi17} and present corresponding, albeit weaker, relationships between $R_{\rm eff}$ and the host and accretion parameters.        

Although we have now identified and quantified the higher dimensional connections between these properties  (Table~\ref{table:regressions}), the origins of these relationships, and particularly the lingering scatter around them, are not yet clear.  Such ambiguity will likely remain until larger samples with considerably better angular resolution and sensitivity are available.  There are simply too many forking paths to generalize with the limited information at hand, since the detailed distribution of the continuum emission may indeed provide the hidden links to the demographic connections studied here.  Nevertheless, we hope to illuminate some broadly relevant issues by exploring these parameter relationships in highly simplified scenarios.  With that more modest goal in mind, we  will explore two limiting approximations.  

First, we need to design a population model for the host properties in this sample to enable a more general interpretation of their connections to the disk parameters.  Figure~\ref{fig:stars_covar} summarizes that model characterization, showing probability densities in grayscale with the mean behaviors in red, and compares it to the individual measurements.  This host population model was manually designed and tuned as described in Appendix~\ref{appendixB}.  Briefly, it employs a three-part linear model of (visual, though not necessarily physical) sub-populations in \{$\log{T_{\rm eff}}$, $\log{\tau_\ast}$\}-space (bottom left corner of Figure~\ref{fig:stars_covar}).  

And second, we make two assumptions that will enable a more intuitive (both physically and mathematically) exploration.  The first is that the continuum emission distribution has a distinct outer boundary, $R_o$.  This rigidity compared to the surface brightness models used in Section~\ref{sec:analysis} is solely designed to simplify the interpretations and focus on the basic results.  In the parlance of the ``Nuker" profile models, it is worth noting that the vast majority of the disks in this sample have $\lesssim$20\%\ of their total emission originating outside $R_t$ (i.e., crudely approximated here as $R_o$), due to their relatively large $\beta$ indices.  We will briefly touch on the effects of ignoring this component of the emission profile below.  The second assumption is that the radial dust temperature profile has a simple parameterization,      
\begin{equation}
    T_d(r) = T_0 \, \left(\frac{L_\ast}{L_\odot}\right)^{0.25} \, \left(\frac{r}{r_0}\right)^{-q} \, ,
    \label{eq:tdust}
\end{equation}
with a fiducial normalization of $T_0 = 30$\,K at $r_0 = 10$\,au for $L_\ast = L_\odot$.  This behavior is motivated by the (admittedly coarse) radiative transfer grid calculated by \citet{andrews13}, and is roughly consistent with measurements that locate the CO condensation fronts at $\sim$20\,K in the disks around two benchmark sample members, TW Hya \citep{qi13} and HD 163296 \citep{qi15} for reasonable values of $q$.

\subsection{Optically Thick Scenario} \label{subsec:thick}

The first scenario we will consider is the case where all of the emission is optically thick ($\tau_d \gg 1$).  We define a corresponding brightness profile,
\begin{equation}
    I_\nu(r) \approx \begin{cases}
    \mathcal{F} \, B_\nu(T_d) \, & \, \text{if $r \le R_o$} \, .\\
    0 \,  & \, \text{otherwise} \, ,
  \end{cases}
\end{equation}
where $\mathcal{F}$ is a constant, tunable, {\it intensity-weighted} ``filling factor" ($0 < \mathcal{F} \le 1$).  To simplify the discussion, let us temporarily assume the disks in this sample can be described in the Rayleigh-Jeans approximation, so that
\begin{equation}
    I_\nu(r) \approx \frac{2 \, \nu^2 \, k}{c^2} \, \mathcal{F} \, T_0 \, r_0^q \, \left(\frac{L_\ast}{L_\odot}\right)^{0.25} \, r^{-q}
\end{equation}
inside $R_o$.  Then, following Equation~\ref{eq:fcum}, the cumulative intensity profile is
\begin{equation}
    f_\nu(r) = \frac{4 \pi \, \nu^2 \, k}{(2-q) \, c^2} \, \mathcal{F} \, T_0 \, r_0^q \, \left(\frac{L_\ast}{L_\odot}\right)^{0.25} \, r^{2-q} \, ,
\end{equation}
and, recalling the definitions of the key observables $L_{\rm mm} = f_\nu(R_o)$ and $x L_{\rm mm} = f_\nu(R_{\rm eff})$, we can relate the continuum luminosity and effective size to the model,
\begin{subequations}
    \begin{equation}
        L_{\rm mm} = \frac{4 \pi \, \nu^2 \, k}{(2-q) \, c^2} \, \mathcal{F} \, T_0 \, r_0^q \, \left(\frac{L_\ast}{L_\odot}\right)^{0.25} \, R_o^{2-q} \, ,
        \label{eq:Lmm_Rt_thick}
    \end{equation}
    \begin{equation}
        R_{\rm eff} = x^{1/(2-q)} \, R_o \, .
        \label{eq:Reff_Rt_thick}
    \end{equation}
    \label{eq:base_thick}
\end{subequations}
These are the ``base" relations that connect the disk observables \{$L_{\rm mm}$, $R_{\rm eff}$\} to the stellar population (through $L_\ast$, via the population model shown in Figure~\ref{fig:stars_covar}) and the model parameters \{$q$, $\mathcal{F}$, $R_o$\}.  

\begin{figure*}[t!]
\includegraphics[width=\linewidth]{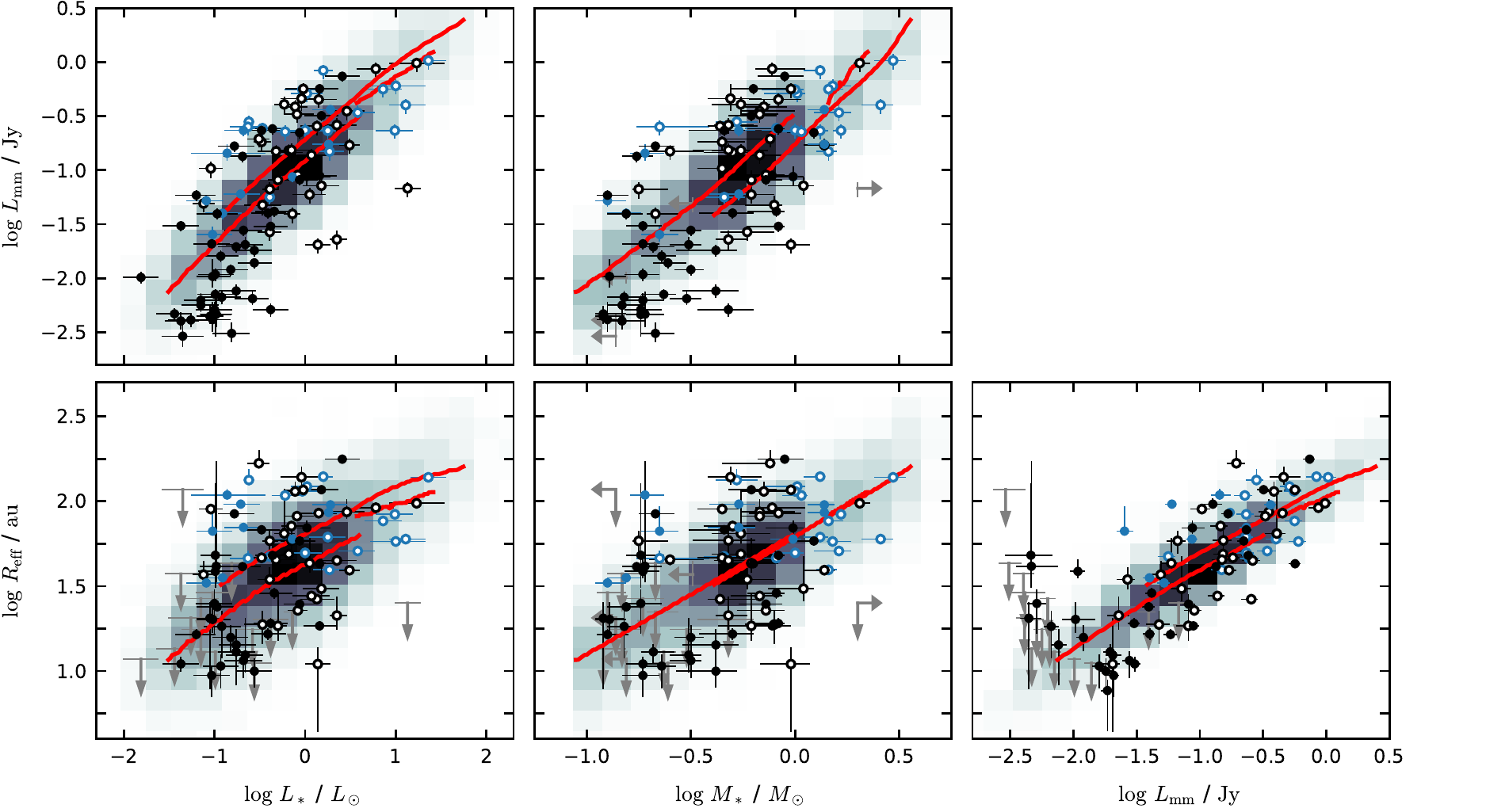}
\figcaption{The relationships between disk continuum and stellar host parameters for the joint sample (points, symbols as in Figure~\ref{fig:hrd}.  The red curves describe the mean behavior of a population model that assumes all of the emission is optically thick, with the assumptions described in Equation~\ref{eq:thickmodel}.  The grayscale is a probability density map for that mean population model which folds in some appropriate scatter in $T_0$, $q$, $\mathcal{F}$, and the prescribed $R_o$--$M_\ast$ scaling, along with the underlying stellar population model.  
\label{fig:thick_covar}
}
\end{figure*}

Combining Equation~\ref{eq:base_thick} and re-arranging terms, we find
\begin{equation}
    R_{\rm eff} \propto L_\ast^{-1/4(2-q)} \, L_{\rm mm}^{1/(2-q)} \, .
    \label{eq:thick_link}
\end{equation}
Therefore, in this scenario the {\it shapes} of both the $R_{\rm eff}$--$L_{\rm mm}$ and $L_\ast$--$L_{\rm mm}$ scaling relations derived in Section~\ref{subsec:scalings} (see Table~\ref{table:regressions}) are reproduced when $q = 0.57\pm0.08$, a reasonable temperature gradient for irradiated accretion disks \citep[e.g.,][]{dalessio98}.  Going back to the base relations in Equation~\ref{eq:base_thick}, we can re-cast the requirement to reproduce these scaling relations in terms of the model parameters and find that $R_o \propto L_\ast^{0.4}$, or, for the mean $M_\ast$--$L_\ast$ relationship, $R_o \propto M_\ast^{0.7}$.    

The {\it normalization} of that $R_o$--$M_\ast$ relation needs to be chosen to reproduce the observed $R_{\rm eff}$--$M_\ast$ behavior (or its equivalents; see Figure~\ref{fig:lmmreff_hosts}).  However, for the fiducial $T_d$ normalization and a continuous distribution of emitting dust ($\mathcal{F} = 1$), this model scenario generates too much emission (i.e., $L_{\rm mm}$ is too high for any given $L_\ast$, $M_\ast$, or $R_{\rm eff}$).  The magnitude of the discrepancy is about a factor of six in $L_{\rm mm}$, although roughly half of that is a result of presuming the Rayleigh-Jeans approximation is valid; a more accurate discrepancy is a factor of three.  The only ways to reconcile this offset are to make the disks colder (e.g., decrease $T_0$ from $\sim$30 to 10\,K for solar luminosity hosts), to selectively remove some emitting material (i.e., set $\mathcal{F} \approx 0.3$), or some combination of those options.  The first option alone is difficult to achieve for realistic radiation transfer calculations, so we prefer the second.  There are many ways to distribute the emitting material to fulfill that criterion, but from the demographics perspective considered here the arrangement itself is not important.   

Figure~\ref{fig:thick_covar} compares the observed relationships between \{$L_{\rm mm}$, $R_{\rm eff}$, $L_\ast$, $M_\ast$\} for the joint sample with a population model of the optically thick scenario described above (using the full Planck function instead of the Rayleigh-Jeans approximation).  The mean behavior in this multi-dimensional parameter-space can be explained well with three key assumptions:
\begin{equation}
    \begin{cases}
        \,\, q \approx 0.57 \, , \\
        \,\, R_o \approx 90 \, (M_\ast/M_\odot)^{0.7} \, \text{au} \, , \\
        \,\, \mathcal{F} \approx 0.3 \, .
    \end{cases}
    \label{eq:thickmodel}
\end{equation}
The scatter around these mean relationships has modest contribution ($\sim$30--50\%) from the dispersion in the stellar host parameters.  Neither those variations, nor the additional scatter imposed by reasonable uncertainty in $q$ ($\pm$0.08) or even in $T_0$ ($\pm$5\,K), nor any variations in $\mathcal{F}$, provide an explanation for the broad distribution observed in the $M_\ast$--$R_{\rm eff}$ plane.  That scatter must be explained in the assumed $R_o$--$M_\ast$ behavior: we find that {\it all} of the observed scatter can be explained if its normalization has a log-normal variation around the mean with a standard deviation of 0.2\,dex and its index has a normal distribution with a standard deviation of 0.1.  The permissible range of $\mathcal{F}$ is $\sim$0.1--0.5 before the model scatter becomes considerably broader than is observed.

\subsection{Optically Thin Scenario} \label{subsec:thin}

\begin{figure*}[t!]
\includegraphics[width=\linewidth]{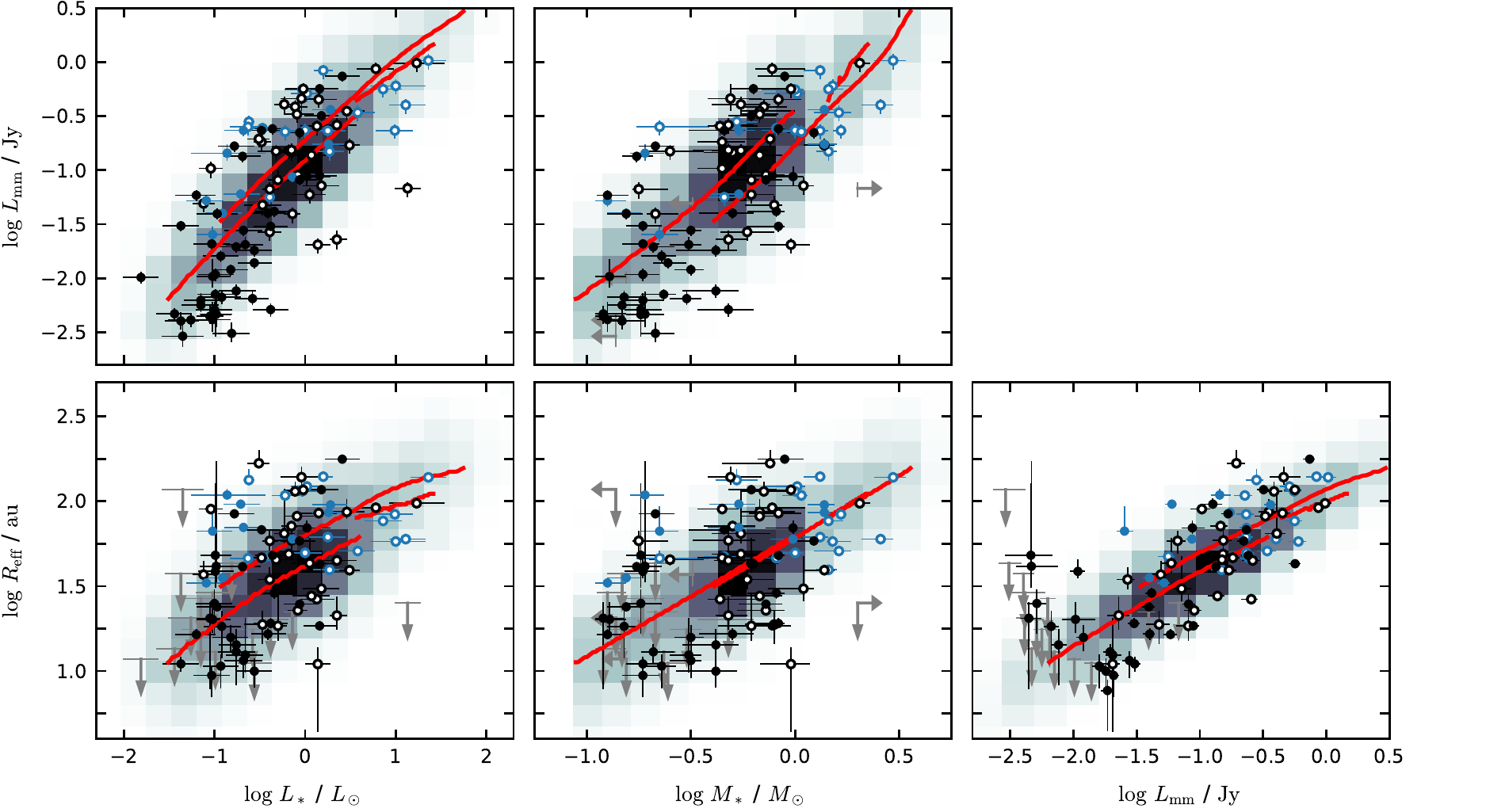}
\figcaption{The same as Figure~\ref{fig:thick_covar}, but for an optically thin disk population model.
\label{fig:thin_covar}
}
\end{figure*}

We also consider the opposite limiting scenario, where the emission is optically thin ($\tau_d \ll 1$) and
\begin{equation}
    I_\nu(r) \approx \begin{cases}
    B_\nu(T_d) \, \tau_d & \, \text{if $r \le R_o$} \, .\\
    0 \,  & \, \text{otherwise} \, .
  \end{cases}
\end{equation}
This scenario is mathematically similar to the optically thick case, but with the added complexity of permissible variations in $\tau_d$.  We assume the generic behavior
\begin{equation}
    \tau_d(r) = \tau_0 \, \left(\frac{L_\ast}{L_\odot}\right)^\eta \, \left(\frac{r}{r_0}\right)^{-p} \, .
    \label{eq:tau}
\end{equation}
To again analytically illustrate the scaling behaviors in this model, we (temporarily) assume that the Rayleigh-Jeans approximation is applicable, so
\begin{equation}
    I_\nu(r) \approx \frac{2 \, \nu^2 \, k}{c^2} \, T_0 \, \tau_0 \, r_0^{p+q} \, \left(\frac{L_\ast}{L_\odot}\right)^{0.25+\eta} \, r^{-(p+q)}
\end{equation}
inside $R_o$.  Integrating that brightness profile and relating it to the definitions of the continuum luminosity and effective size, we arrive at the ``base" relations, 
\begin{subequations}
    \begin{equation}
        L_{\rm mm} = \frac{4 \pi \nu^2 k T_0 \, \tau_0 \, r_0^{p+q}}{(2-p-q) \, c^2} \, \left(\frac{L_\ast}{L_\odot}\right)^{0.25+\eta}  R_o^{2-p-q} \, ,
        \label{eq:Lmm_Rt_thin}
    \end{equation}
    \begin{equation}
        R_{\rm eff} = x^{1/(2-p-q)} \, R_o  
        \label{eq:Reff_Rt_thin}
    \end{equation}
    \label{eq:base_thin}
\end{subequations}
(cf., Equation~\ref{eq:base_thick} for the $\tau_d \gg 1$ case).

Combining the parts of Equation~\ref{eq:base_thin}, we find the somewhat unwieldy scaling behavior
\begin{equation}
    R_{\rm eff} \propto L_\ast^{-(\eta+0.25)/(2-p-q)} \, L_{\rm mm}^{1/(2-p-q)} \, .
\end{equation}
In this case, the criterion required to simultaneously reproduce the observed shapes of the $R_{\rm eff}$--$L_{\rm mm}$ and $L_\ast$--$L_{\rm mm}$ scaling relations is
\begin{equation}
    \eta \approx 0.4(p+q) - 0.2 \, .
\end{equation}
For reasonable $p+q$ (based on the $\gamma$ values inferred in Section~\ref{sec:results}), that again implies that $R_o \propto L_\ast^{0.4}$, or equivalently $R_o \propto M_\ast^{0.7}$ in this case (as it must to explain the observed $R_{\rm eff}$--$L_\ast$ or --$M_\ast$ behaviors).  

Figure~\ref{fig:thin_covar} shows the parameter relationships of interest overlaid with optically thin population models that were manually tuned to match the data.  In those model calculations, we assumed the more general surface brightness prescription $I_\nu \approx B_\nu(T_d) (1 - e^{-\tau_d})$, although the simplification described above still provides a reasonable approximation.  As was the case for the optically thick models, we need to make three assumptions to reproduce the trends, relating to a gradient, a size-scaling, and a normalization,
\begin{equation}
    \begin{cases}
        \,\, \eta \approx 0.4(p+q) - 0.2 \, , \\
        \,\, R_o \approx 90 \, (M_\ast/M_\odot)^{0.7} \, \text{au} \, , \\
        \,\, \tau_0 \approx 0.4 \, (M_\ast/M_\odot)^{1.7 \eta} \, ,
    \end{cases}
    \label{eq:thinmodel}
\end{equation}
where in the last criterion we substituted the mass scaling for the original luminosity scaling using the mean $L_\ast$--$M_\ast$ relation appropriate for this sample.  In Figure~\ref{fig:thin_covar}, we have assumed $p+q \approx 0.75\pm0.25$ ($\eta \approx 0.10\pm0.11$) to be consistent with the measured surface brightness slopes (note that $p+q \approx \gamma$ in this model).  

\begin{figure*}[t!]
\includegraphics[width=\linewidth]{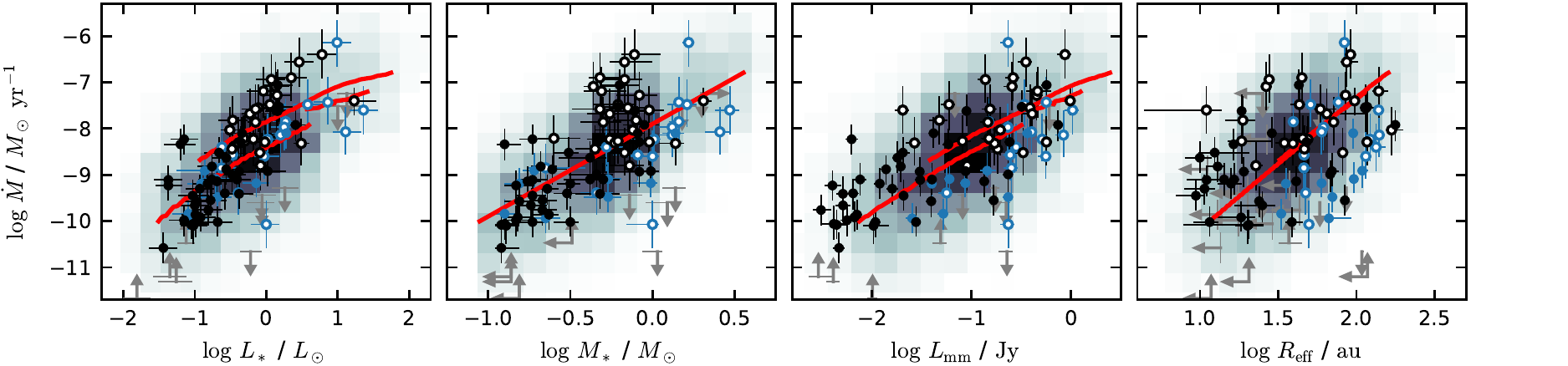}
\figcaption{An analogous comparison as Figure~\ref{fig:thick_covar}, but in this case for the added dimension of accretion rate ($\dot{M}_\ast$).  These behaviors presume the same stellar population model, the optically thick prescription from Section~\ref{subsec:thick}, and an intrinsic $\dot{M}_\ast$--$M_\ast$ scaling (see Table~\ref{table:regressions}).  
\label{fig:mdot_covar}
}
\end{figure*}

We want to explicitly point out two consequences of this formulation.  First, this behavior, where $\eta$ is roughly zero, is consistent with a scenario where all disks have the {\it same optical depth profile}, regardless of their host properties \citep[cf., Eq.~\ref{eq:tau}; see also][]{pietu14}: the observed scaling relations can all be explained with variations in the disk sizes.  And second, if we make the standard assumption that the dust opacity $\kappa_\nu$ does not vary with radius, we can derive a scaling relation between the disk mass and host mass.  To do that, under these assumptions ($\eta \sim 0$ and $\kappa_\nu(r) \approx {\rm constant}$) we note that the dust surface density profile would vary like
\begin{equation}
    \Sigma \propto M_\ast^{\xi} \, r^{-p},
\end{equation}
where we have permitted an $M_\ast$-dependence in the normalization.  The dust mass is the integral of the surface density profile over the disk area, and so 
\begin{equation}
    M_d = \int_0^{R_o} \Sigma(r) \, 2\pi \, r \, dr \propto M_\ast^\xi \, R_o^{2-p}.
\end{equation}
Since $R_o \propto M_\ast^{0.7}$ (Eq.~\ref{eq:thinmodel}), and $p = 0.25$ for the fiducial case where $q = 0.5$, we find that $M_d \propto M_\ast^{1.2+\xi}$.  The salient point is that a consideration of disk sizes in the demographic landscape shows that the roughly linear $M_d$--$M_\ast$ scaling relation inferred for the typical set of assumptions \citep{andrews13,ansdell16,pascucci16} implies that disks have similar surface density profiles regardless of their host parameters (i.e., $\xi$ is consistent with zero).  It is worthwhile to consider these implications in studies that employ the scaling relations that are inferred with such assumptions.  

In terms of the scatter in these parameter relationships, we are again forced to assume the same dispersion around the mean $R_o$--$M_\ast$ relation as in the optically thick case.  Some additional scatter in the optical depth normalization ($\tau_0$) is also plausible, so long as the standard deviation in a log-normal distribution is $\lesssim$0.4\,dex (and $\sim$0.2\,dex is more appropriate).

\subsection{Links to Accretion Rates} \label{subsec:mdot}

\begin{figure}[t!]
\includegraphics[width=\linewidth]{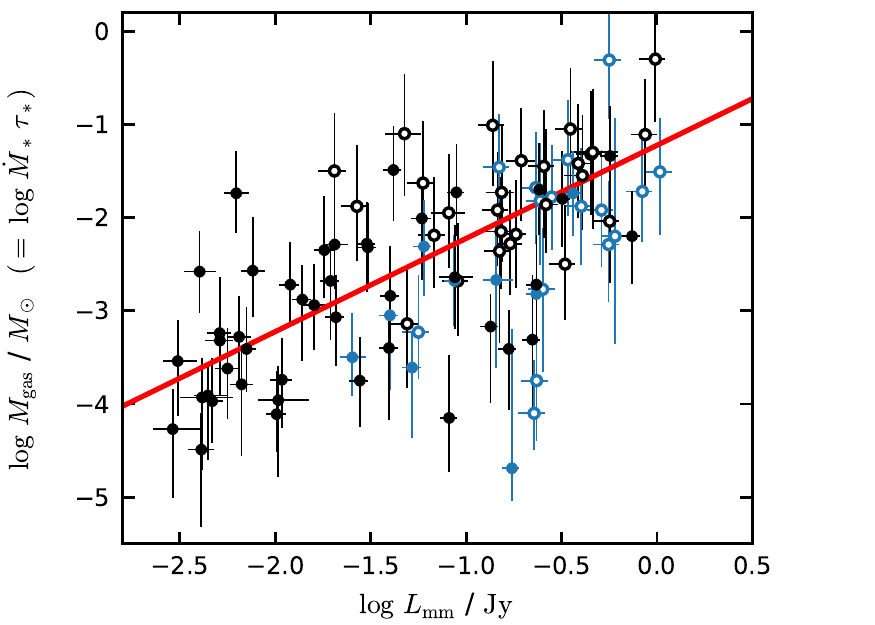}
\figcaption{A comparison of the continuum luminosities ($\propto M_{\rm dust}$) and a simple diagnostic of the gas mass in standard accretion disk models ($M_{\rm gas} \approx \dot{M}_\ast \, \tau_\ast$; cf., \citealt{hartmann98}).  The red curve marks the expected behavior for standard opacity and temperature assumptions if the dust-to-gas mass ratio is 1\%.
\label{fig:mgas}
}
\end{figure}

An example of the extensions of these population models to include the mass accretion rate is shown in Figure~\ref{fig:mdot_covar}, in the case for the optically thick scenario (although the optically thin scenario works just as well).  The added assumption when expanding the dimensionality this way is that there is an intrinsic $\dot{M}_\ast$--$M_\ast$ relation like the one observed (including the scatter; see Table~\ref{table:regressions}).  We could just as easily have presumed an intrinsic $L_{\rm mm}$--$\dot{M}_\ast$ relationship as observed \citep[e.g., see][]{mulders17}, and thereby would predict an appropriate $\dot{M}_\ast$--$M_\ast$ scaling and amount of scatter.  Figure~\ref{fig:mgas} makes the comparison between $L_{\rm mm}$ and a crude diagnostic of the gas mass of the disk, $M_{\rm gas} \approx \dot{M}_\ast \, \tau_\ast$ \citep[cf.,][]{hartmann98}.  \dt{The observed correlation is to be expected, since $\dot{M}_\ast$ and $L_{\rm mm}$ are correlated (Figure~\ref{fig:lmmreff_acc}) and neither of those variables depend significantly on $\tau_\ast$.  For the assumptions outlined above, the same behavior would be produced for the optically thick or thin population models.  That said, the latter models may seem more compelling in this context} because $L_{\rm mm}$ is roughly proportional to the dust mass ($M_{\rm dust}$).  The red curve in Figure~\ref{fig:mgas} is not a fit: it represents the \dt{expected behavior for equivalent masses if the disks are optically thin,} have a dust-to-gas mass ratio of 1\%\, a mean dust temperature of 20\,K, and a 340\,GHz dust opacity of 3.5\,cm$^2$\,g$^{-1}$.  

It is tempting to fold the added dimensionality of {\it size} into an interpretation in the context of the viscous evolution of simple accretion disks \citep[e.g.,][]{hartmann98,andrews09,isella09,pascucci16,rafikov17,tazzari17,lodato17,mulders17}.  The effective continuum sizes and luminosities as measured here are not easily associated with the viscous model parameters of interest (the gas mass and characteristic radius) because they are both strongly (and non-linearly) modulated by the coupled evolution of the gas and solids in the disk \citep[e.g.,][]{birnstiel12,birnstiel15}.  Moreover, models that perform that coupling are known to have a significant efficiency problem \citep[e.g.,][]{takeuchi02,takeuchi05,brauer07}, presumably requiring disk structures that deviate significantly from the standard gas model configurations \citep[e.g.,][]{pinilla12a}.  For those reasons, our preference is to keep the focus on more empirical relationships.  Nevertheless, it would be interesting in future work to consider full population models that link the evolution of solids in viscous accretion disks in an effort to more accurately assess their ability (or not) to reproduce the observed scalings on the right timescales.  

\subsection{Comments on Transition Disks} \label{subsec:trans}

To this point, we have discussed the multidimensional relationships among all of the sample targets together, regardless of their continuum emission morphology.  The sub-sample of transition disks (see Section~\ref{subsec:results} for our definition) have bulk demographic properties that, broadly speaking, match well with the overall sample.  But aside from that general agreement, there are some subtle deviations that merit brief attention.  

First, and perhaps most obviously, the transition disks in this sample tend to be hosted by more luminous (massive) stars.  We presume that this is a selection effect, in that there could well be other transition disks among the less luminous (massive) members of the sample that have not yet been identified \citep[e.g., see][]{ercolano09}.  However, this could instead be a real physical effect; very high angular resolution imaging would be required to know for certain.  Figure~\ref{fig:stars_covar} suggests that the transition disk hosts are slightly less luminous than their ``normal" counterparts at a fixed effective temperature, and therefore appear older.  Their inferred age distribution does indeed peak at 2.5\,Myr, compared to 1.5\,Myr for the normal disks, but a shift of that amount is not statistically significant given the large uncertainties.  

\citet{pinilla18} recently argued that the $M_d$--$M_\ast$ (or equivalently $L_{\rm mm}$--$M_\ast$) scaling for transition disks is considerably flatter than for the normal disk population.  A close examination of the top right panel of Figure~\ref{fig:lmmreff_hosts} indeed shows a similar result for this sample.\footnote{There is substantial sample overlap between this article and the \citet{pinilla18} study, particularly at the low-$M_\ast$ end.}  The normal disk relation is the same as quoted in Table~\ref{table:regressions}, but the transition disks alone have a much lower scaling index ($\mathcal{B} = 0.86\pm0.18$; a difference at the 3\,$\sigma$ level).  We should caution, however, that both this analysis and the \citet{pinilla18} result are strongly influenced by the 2 or 3 targets at the low-$M_\ast$ end.  If those are just the brightest transition disks with low-mass hosts, and there are indeed more such systems with fainter disks that we have not yet identified, this apparent discrepancy could easily be biased by selection effects.  We also find marginal support for the claim by \citet{najita07,najita15} that the transition disks have preferentially lower accretion rates for a given disk mass (see Figures~\ref{fig:lmmreff_acc} or \ref{fig:mdot_covar}).  A comparison of the regression results in the $L_{\rm mm}$--$M_\ast$ plane finds that the normalization (intercept) is $\sim$0.6\,dex lower for transition disks compared to normal disks (at $\sim$90\%\ confidence); both sub-samples have the same slopes.          

Finally, we note that the surface brightnesses measured in Section~\ref{sec:analysis} for the transition disks are high.  Assuming the standard dust temperature prescription (Equation~\ref{eq:tdust}), we would infer that all of these disks are optically thick around their peaks.  One could reasonably argue that the local $T_d$ at these peaks should be enhanced by direct stellar irradiation \citep[e.g.,][]{dullemond01,dalessio05}, but not to a level that would substantially reduce the optical depth estimates.  In reality, the ``ring" features for these systems are not resolved well, so if they are more narrow than we can infer then the peak brightness temperatures (and thereby optical depths) could actually be higher.  This is all an interesting manifestation of the first scenario to explain the multi-dimensional scaling relations, described in Section~\ref{subsec:thick}: the emission is optically thick and has a large effective size, but the depleted central cavities reduce the continuum luminosity to a level consistent with a ``filling factor" ($\mathcal{F}$) of $\sim$0.1--0.5.  Since the transition disks generally have similar demographic behaviors to the general population, it is natural to wonder if they are just a more obvious manifestation of this kind of behavior.  Perhaps this is a clue that the multidimensional relationships probed here are produced by optically thick substructures on spatial scales considerably smaller than the available resolution.

\subsection{Further Caveats and Future Work} \label{subsec:hybrid}

Of course, the limiting approximations in Sections~\ref{subsec:thick} and \ref{subsec:thin} are vast over-simplifications.  The reality is probably a messier hybrid of those scenarios, perhaps where there are higher-level dependencies between the model parameters and the host and/or disk properties.  In a sense, this could still be treated in the optically thick substructures paradigm if we permit a complicated functional form for the ``filling factor" $\mathcal{F}$.  That parameter can include optically thin contributions, depleted zones, or both, in myriad morphological configurations.  

\begin{figure}[t!]
\includegraphics[width=\linewidth]{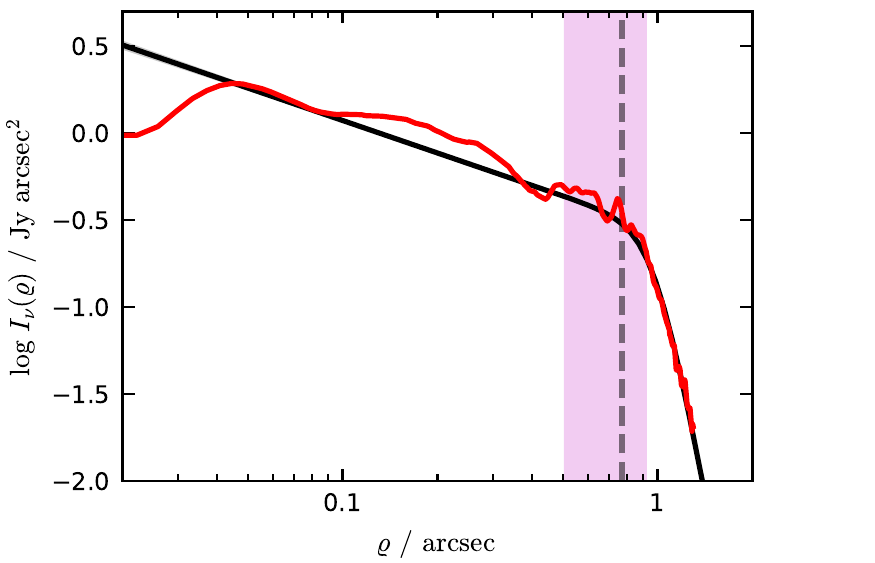}
\figcaption{The 340\,GHz continuum surface brightness distribution model inferred from the 0\farcs3 SMA observations of the TW Hya disk (black), overlaid with the observed (deprojected) ALMA emission profile at 20\,mas resolution \citep{andrews16}.  The vertical dashed line marks the $\varrho_{\rm eff}$ for this target listed in Table~\ref{table:SBpars2}.  The shaded region marks the radial separations where the emission transitions from optically thick to thin, based on the brightness temperatures and spatially resolved spectral index measurements of \citet{tsukagoshi16} and \citet{huang18}.  
\label{fig:twhya}
}
\end{figure}

This may seem like a good explanation of demographic trends will rely on some knowledge of complicated minutiae.  In reality, that is probably the case: it will be difficult to distinguish how to properly formulate population models without observations at much higher resolution.  However, we can start to check if even the simple pictures laid out above make sense for specific examples.  Perhaps the most obvious case to illustrate the point is the TW Hya disk, which has been observed at very high angular resolution (20--30\,mas; $\sim$2\,au) with ALMA (Andrews et al.~2016).  Figure~\ref{fig:twhya} shows the model brightness profile inferred from the more modest-resolution SMA observations \citep{andrews16,tripathi17}, overlaid with the ALMA data.  The spectral index of this emission is roughly constant at $\approx$2 inside $\sim$0\farcs5 (and perhaps out to 0\farcs9); coupled with the high brightness temperatures (comparable to the expected $T_d$), it is clear that the inner disk emission is optically thick \citep{tsukagoshi16,huang18}.  In this case, the optically thick filling factor is $\mathcal{F} \approx 0.4$--0.6; if we further account for the small-scale radial depletions (``gaps"), $\mathcal{F}$ decreases by another $\sim$0.1.  So, in this case, a high-resolution knowledge of where the emission is optically thick lends some credence to an explanation  similar to the scenario described in Section~\ref{subsec:thick}.               

Ultimately, similar evidence as for TW Hya would be needed to definitively connect physical models to the demographic properties discussed here (or perhaps more elaborate ones).  That said, some tests of the optically thick scenario might be available by folding in spectral index information (perhaps even unresolved) to the population analyses.  Some guidance from linking models of viscous evolution and the coupling between gas and solids would also be welcome.

\section{Summary} \label{sec:summary}

We have combined the archival SMA measurements from \citet{tripathi17} with the comparable archival ALMA survey in Lupus \citep{ansdell16} to conduct a homogeneous analysis of the resolved 340\,GHz continuum emission from 105 nearby protoplanetary disks, with a focus on the multidimensional demographics related to emission {\it sizes}.  Our key findings are:
\begin{itemize}

\item We confirm (in form and normalization) and quantify the previous measurements of strong correlations between the continuum luminosities ($L_{\rm mm}$) and the stellar host masses ($M_\ast$) or luminosities ($L_\ast$) -- $L_{\rm mm} \propto M_\ast^{1.5}$ or $L_{\rm mm} \propto L_\ast^{0.8}$ -- as well as the accretion rates -- $L_{\rm mm} \propto \dot{M}_\ast^{1.0}$.
  
\item We verify the relationship between the continuum emission size ($R_{\rm eff}$) and $L_{\rm mm}$ measured by \citet{tripathi17}: the amount of emission scales linearly with its surface area: $L_{\rm mm} \propto R_{\rm eff}^{2.0}$.

\item We identify new, albeit weaker, scaling relations connecting the emission sizes and the host properties -- $R_{\rm eff} \propto M_\ast^{0.6}$ or $\propto L_\ast^{0.3}$ -- and a marginal connection with the accretion rates -- $R_{\rm eff} \propto \dot{M}_\ast^{0.9}$.  

\item With some simplified approximations, demographic models explain these scaling relations and their associated dispersions for either optically thick or thin emission.  The thick case requires an intensity-weighted filling factor of $\sim$0.3.  The thin case suggests that disks have a relatively uniform optical depth profile that is independent of the host properties: the distribution of continuum luminosities are explained with variations in the disk size.  In both cases, we require a slightly sub-linear scaling (and considerable dispersion) between the dust disk size and the host mass.  

\item The transition disks appear superficially to exhibit the same behavior as the general population, but may show some subtle differences in their connections to the stellar host and accretion parameters.  Selection effects are still an issue in making firm conclusions in this regard.

\item The key takeaway point is that the standard demographic analyses show clear connections among the disk masses ($\sim$$L_{\rm mm}$), disk sizes, and host properties.  Unambiguously disentangling those connections may require large mm continuum surveys at a few times better angular resolution, preferably with spectral index information, to assess the role of optical depth in shaping the observables.

\end{itemize}

\acknowledgments We appreciate useful conversations with Leonardo Testi, Stefano Facchini, Eugene Chiang, and Til Birnstiel.  MT is grateful for support from the Smithsonian Astrophysical Observatory Latino Initiative Program.  The Submillimeter Array is a joint project between the Smithsonian Astrophysical Observatory and the Academia Sinica Institute of Astronomy and Astrophysics and is funded by the Smithsonian Institution and the Academia Sinica.  This paper makes use of the following ALMA data: ADS/JAO.ALMA \#2013.1.00220.S, ADS/JAO.ALMA \#2013.1.00694.S, ADS/JAO.ALMA \#2012.1.00761.S, and ADS/JAO.ALMA \#2013.1.00374.S. ALMA is a partnership of ESO (representing its member states), NSF (USA) and NINS (Japan), together with NRC (Canada), MOST and ASIAA (Taiwan), and KASI (Republic of Korea), in cooperation with the Republic of Chile. The Joint ALMA Observatory is operated by ESO, AUI/NRAO and NAOJ.  This work presents results from the European Space Agency (ESA) space mission {\it Gaia}. {\it Gaia} data are being processed by the {\it Gaia} Data Processing and Analysis Consortium (DPAC). Funding for the DPAC is provided by national institutions, in particular the institutions participating in the {\it Gaia} MultiLateral Agreement (MLA). The {\it Gaia} mission website is \url{https://www.cosmos.esa.int/gaia}. The {\it Gaia} archive website is \url{https://archives.esac.esa.int/gaia}.

\facilities{Submillimeter Array (SMA); Atacama Large Millimeter/submillimeter Array (ALMA)}

\software{{\tt CASA} \citep{mcmullin07}, {\tt Numpy} \citep{numpy}, {\tt Matplotlib} \citep{matplotlib}, {\tt Astropy} \citep{astropy}, {\tt emcee} \citep{foreman-mackey13}, {\tt corner} \citep{corner}, {\tt ScottiePippen} \citep[][\url{https://github.com/iancze/ScottiePippen}]{czekala16}, {\tt linmix} \citep[][\url{https://github.com/jmeyers314/linmix}]{kelly07}}

\bibliography{references}

\clearpage

\appendix 

\section{Detailed Model Results} \label{appendixA} 

\begin{figure*}[b!]
\figurenum{A1}
\includegraphics[width=\textwidth]{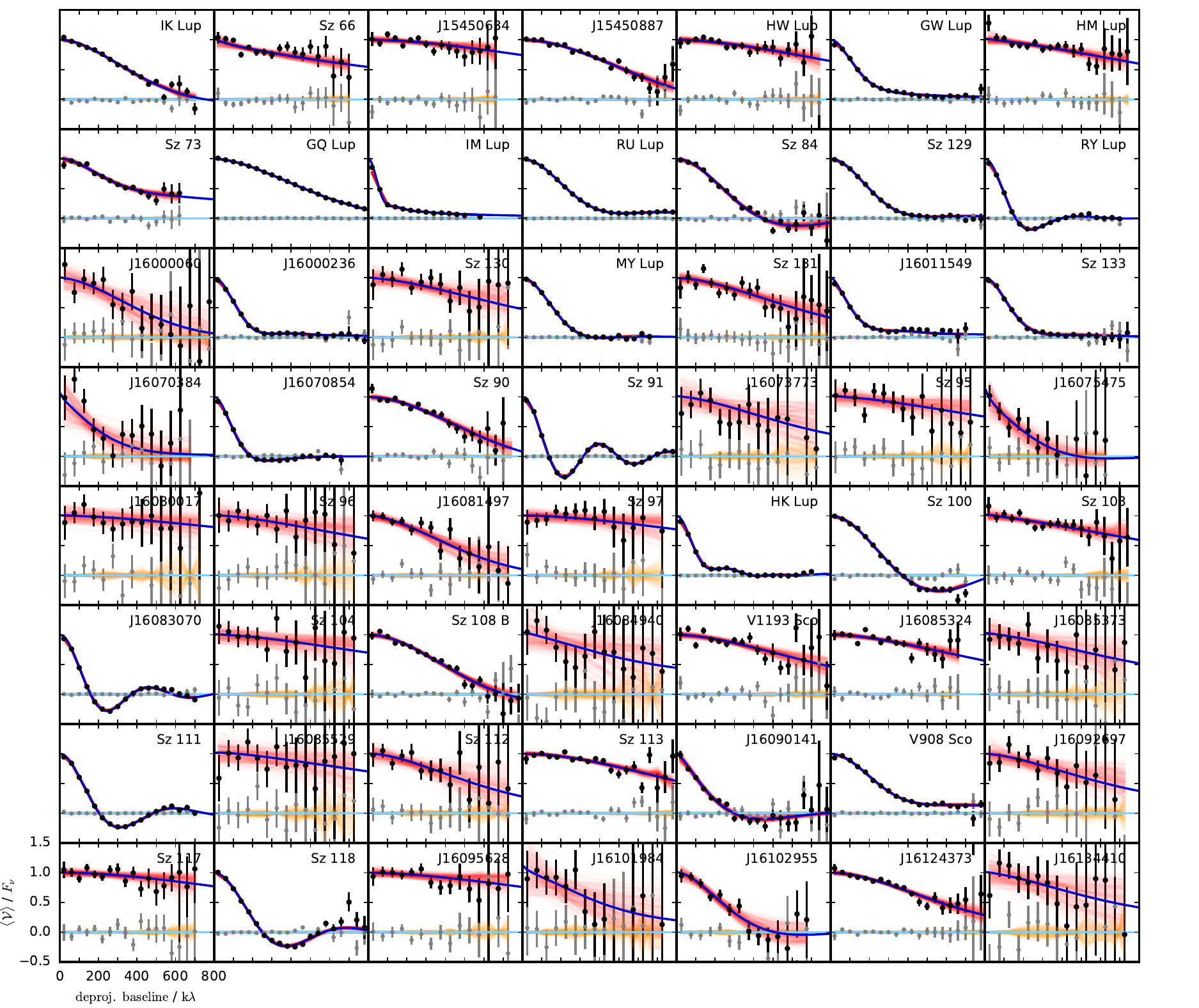}
\figcaption{The azimuthally-averaged, deprojected real (black points) and imaginary (gray points) visibility profiles from each target in the sample, normalized by the median values of the $F_\nu$ posterior distributions.  Red (real) and orange (imaginary) curves show corresponding visibility profiles with the same spatial frequency sampling as the data, constructed from random draws from the model posteriors.  Blue curves represent models with perfect spatial frequency sampling, using an average model of 200 posterior parameter draws.   \label{fig:visgallery}}
\end{figure*}

\begin{figure*}[t!]
\figurenum{A2}
\includegraphics[width=\textwidth]{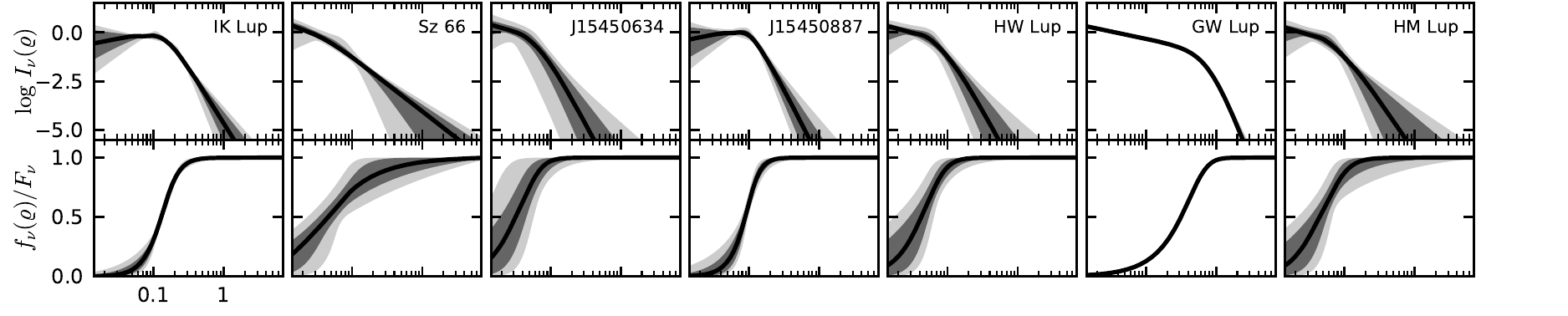}
\includegraphics[width=\textwidth]{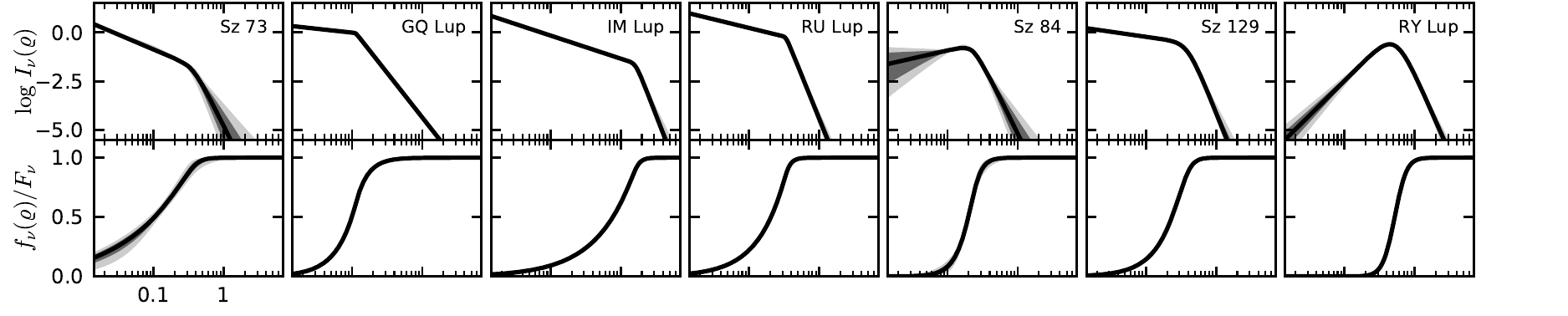}
\includegraphics[width=\textwidth]{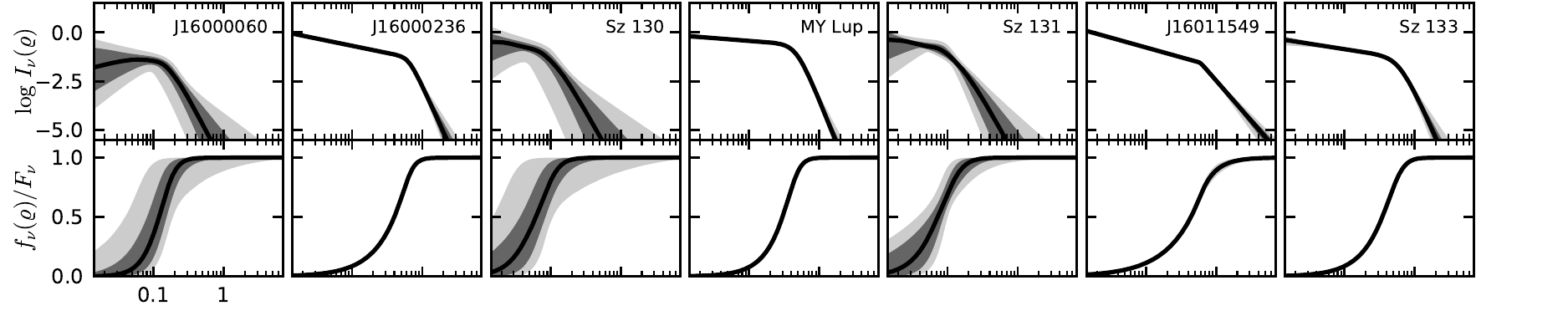}
\includegraphics[width=\textwidth]{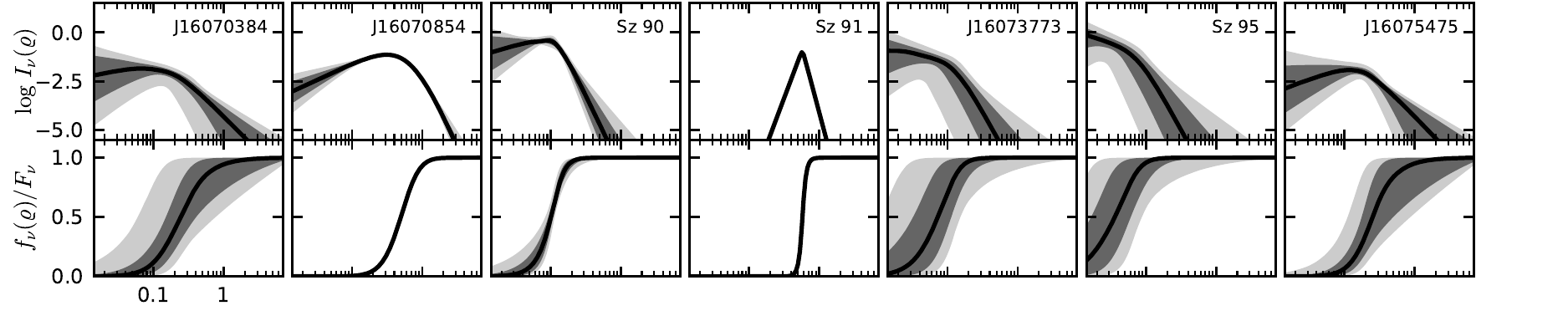}
\includegraphics[width=\textwidth]{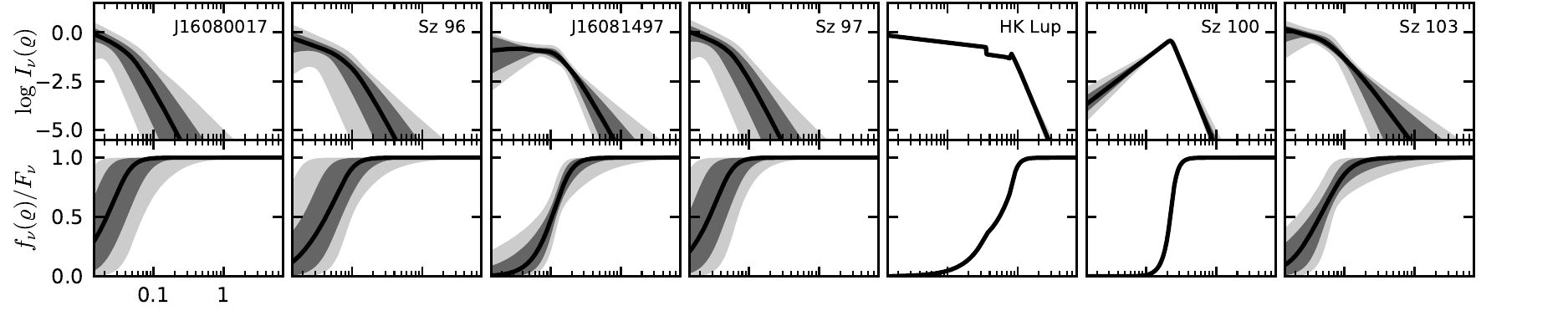}
\includegraphics[width=\textwidth]{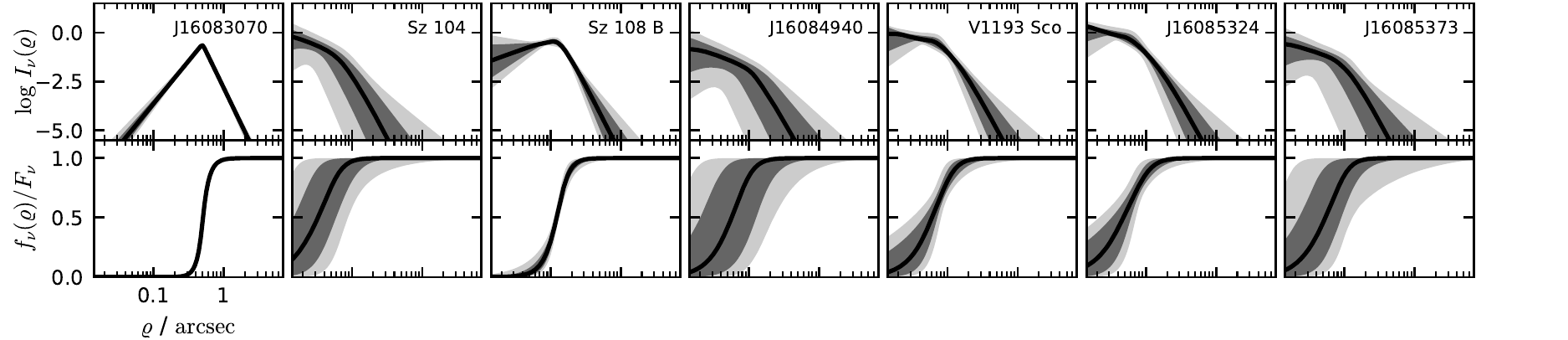}
\figcaption{The surface brightness, $I_{\nu}(\varrho)$, profiles (top panels of each row; in Jy arcsec$^{-2}$ units) and cumulative flux, $f_{\nu}(\varrho)/F_{\nu}$, profiles (bottom panels of each row).  The black curves are the median profiles constructed from the posteriors of the brightness profile model parameters.  The light and dark shaded regions represent the 95 and 68\%\ confidence intervals on those profiles, respectively.   \label{fig:profilesA}}
\end{figure*}

\begin{figure*}[t!]
\figurenum{A3}
\includegraphics[width=\textwidth]{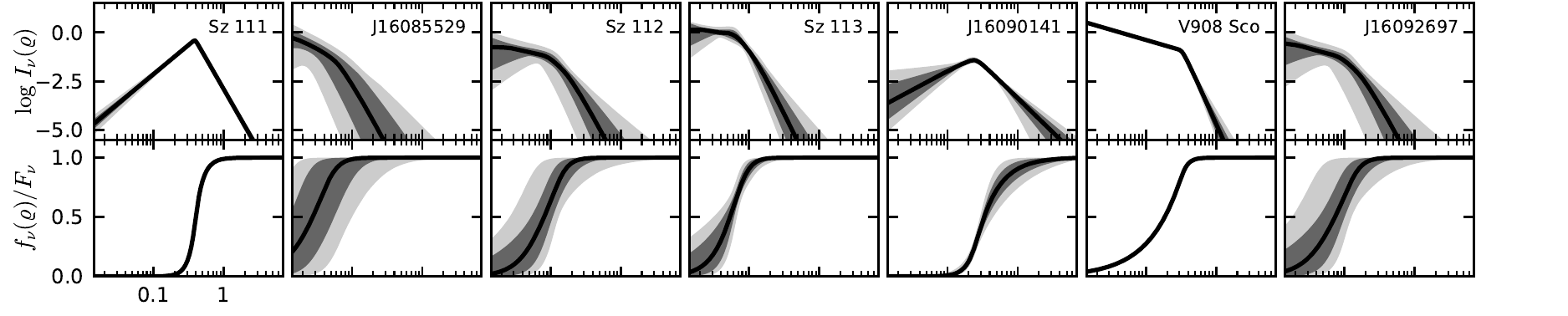}
\includegraphics[width=\textwidth]{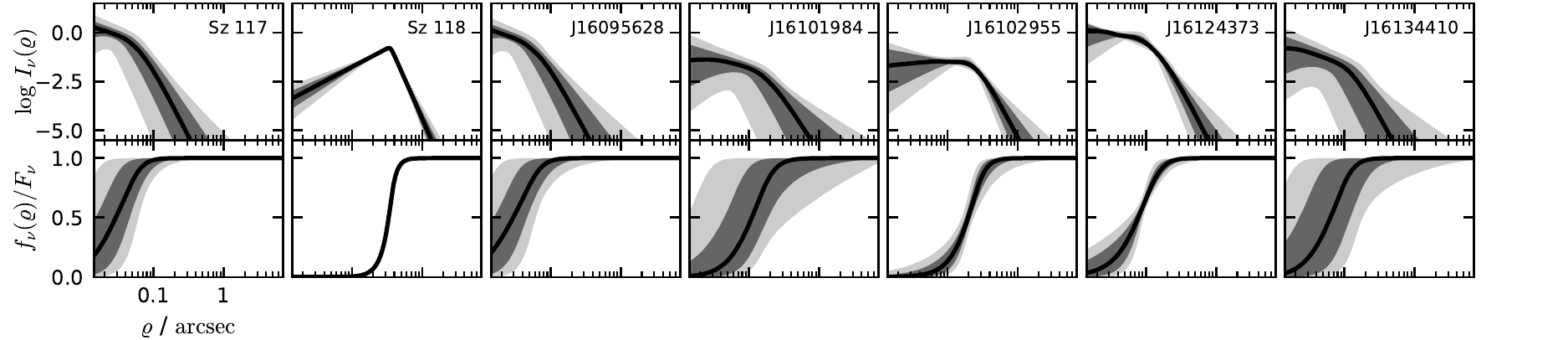}
\figcaption{Surface brightness and cumulative flux radial profiles for additional targets, as in Fig.~\ref{fig:profilesA}.   \label{fig:profilesB}}
\end{figure*}

Figure~\ref{fig:visgallery} shows the deprojected, azimuthally-averaged visibility profiles for each target in the Lupus sample, normalized by their most probable flux densities.  Error bars correspond to the uncertainties on the mean in each spatial frequency bin.  The black and gray points are the real and imaginary components.  The red (orange) curves show the real (imaginary) visibility profiles that correspond to 200 random draws from the surface brightness posteriors, constructed with the same Fourier sampling as the observations and averaged in the same way.  The blue curves are models constructed from the mean of those posterior draws, assuming perfect Fourier coverage.  Figures~\ref{fig:profilesA} and \ref{fig:profilesB} show the confidence intervals on the surface brightness and cumulative intensity profiles for the Lupus sample.  

Table~\ref{table:SBpars} is a compilation of summary statistics for the posterior distributions inferred for the parameters of the continuum surface brightness profiles of the Lupus sample targets.  It includes the effective (angular) sizes, determined from the brightness profiles, and (the logarithms of) that size ($R_{\rm eff}$) and the continuum luminosity ($L_{\rm mm}$) in physical units.  Table~\ref{table:SBpars2} is the equivalent compilation for the SMA sample.  This has been transcribed from Table~3 of the original study by \citet{tripathi17}, but with the last two columns ($R_{\rm eff}$ and $L_{\rm mm}$) adjusted according to the updated distances for each target.  Tables~\ref{table:stars} and \ref{table:starsB} catalog the adopted or inferred values and uncertainties for the distances, stellar properties, and accretion rates for the targets in the Lupus and SMA samples, respectively, along with notes on the literature origins for any relevant assumptions.

\clearpage

\begin{longrotatetable}
\begin{deluxetable}{llccccccc|ccc}
\tabletypesize{\scriptsize}
\tablecaption{Inferred Parameters for the Continuum Surface Brightness Profiles \label{table:SBpars}}
\tablehead{
\colhead{2MASS designation} & \colhead{Name} & \colhead{$F_\nu$ / mJy} & \colhead{$\varrho_t$ / \arcsec} & \colhead{$\log{\alpha}$} & \colhead{$\beta$} & \colhead{$\gamma$} & \colhead{$i$ / \degr} & \colhead{$\varphi$ / \degr} &  \colhead{$\varrho_{\rm eff}$ / \arcsec} & \colhead{$\log{R_{\rm eff}/{\rm au}}$} & \colhead{$\log{L_{\rm mm}/{\rm Jy}}$}
}
\colnumbers
\startdata 
J15392776-3446171  & IK Lup (Sz 65)
                   & \phn66.3 $^{+0.9}_{-0.6}$    & 0.12 $^{+0.08}_{-0.01}$  & $p(\log{\alpha}); \downarrow$
                   & \phn4.5 $^{+4.0}_{-0.1}$     & \phd0.7 $^{+0.1}_{-3.0}$ & 63 $^{+1\phn}_{-1\phn}$
                   & 107 $^{+1\phn}_{-1\phn}$     & 0.16 $^{+0.01}_{-0.01}$  
                   & 1.39 $^{+0.02}_{-0.02}$      & -1.09 $^{+0.04}_{-0.05}$ \\
J15392828-3446180  & Sz 66              
                   & \phn15.1 $^{+3.4}_{-0.2}$    & $<0.17$                  & $p(\log{\alpha}); \downarrow$
                   & \phn2.5 $^{+2.7}_{-0.1}$     & $p(\gamma); \uparrow$    & 69 $^{+12}_{-19}$
                   & \phn51 $^{+15}_{-12}$        & 0.08 $^{+0.05}_{-0.02}$
                   & 1.09 $^{+0.22}_{-0.10}$      & -1.69 $^{+0.07}_{-0.06}$ \\
J15450634-3417378  & J15450634
                   & \phn15.1 $^{+0.6}_{-0.3}$    & 0.04 $^{+0.05}_{-0.02}$  & $p(\log{\alpha}); \downarrow$
                   & $>3.4$                       & $p(\gamma); \uparrow$    & 50 $^{+24}_{-23}$ 
                   & \phn58 $^{+88}_{-31}$        & 0.10 $^{+0.04}_{-0.01}$
                   & 0.89 $^{+0.07}_{-0.24}$      & -1.73 $^{+0.04}_{-0.05}$ \\
J15450887-3417333  & J15450887
                   & \phn47.7 $^{+0.7}_{-0.6}$    & 0.10 $^{+0.04}_{-0.01}$  & $p(\log{\alpha}); \uparrow$   
                   & $>4.4$                       & $p(\gamma); \uparrow$    & 47 $^{+3\phn}_{-6\phn}$ 
                   & 162 $^{+4\phn}_{-7\phn}$     & 0.11 $^{+0.01}_{-0.01}$  
                   & 1.22 $^{+0.03}_{-0.02}$      & -1.23 $^{+0.05}_{-0.06}$ \\
\enddata
\tablecomments{The quoted parameter values are the peaks of their posterior distributions; their corresponding uncertainties represent the 68\%\ confidence interval.  Limits represent the 95\%\ confidence boundary.  A note of ``$p(X)$" means the posterior is consistent with the prior at $<$68\%\ confidence; ``$p(X); \downarrow$"  or ``$p(X); \uparrow$" means the posterior is consistent with the prior at 95\%\ confidence, but has a marginal preference toward the lower or upper bound of the prior on $X$, respectively.  Note that the $F_\nu$ posterior does not include systematic flux calibration uncertainties (although the $\log{L_{\rm mm}}$ values do) and that the parameters to the right of the vertical bar ($\varrho_{\rm eff}$, $\log{R_{\rm eff}}$, and $\log{L_{\rm mm}}$) are not {\it directly} inferred, but rather constructed from the joint posterior on \{$F_{\nu}$, $\varrho_t$, $\log{\alpha}$, $\beta$, $\gamma$\}; see Sect.~\ref{sec:analysis} and \citet{tripathi17} for details.  To facilitate easy comparisons, the (log) luminosities in Col.~(12) are in units of Jy for a distance of 140\,pc.}
\end{deluxetable}
\end{longrotatetable}

\clearpage
\begin{longrotatetable}
\begin{deluxetable}{llccccccc|ccc}
\tabletypesize{\scriptsize}
\tablecaption{Inferred Parameters for the Continuum Surface Brightness Profiles \label{table:SBpars2}}
\tablehead{
\colhead{2MASS designation} & \colhead{Name} & \colhead{$F_\nu$ / mJy} & \colhead{$\varrho_t$ / \arcsec} & \colhead{$\log{\alpha}$} & \colhead{$\beta$} & \colhead{$\gamma$} & \colhead{$i$ / \degr} & \colhead{$\varphi$ / \degr} &  \colhead{$\varrho_{\rm eff}$ / \arcsec} & \colhead{$\log{R_{\rm eff}/{\rm au}}$} & \colhead{$\log{L_{\rm mm}/{\rm Jy}}$}
}
\colnumbers
\startdata 
J03454828+3224118  & LkH$\alpha$ 330   
                   & 0.208 $^{+0.004}_{-0.002}$   & 0.38 $^{+0.45}_{-0.26}$  & 0.83 $^{+0.90}_{-0.10}$
                   & 5.2 $^{+2.7}_{-0.3}$         & -1.7 $^{+0.3}_{-1.0}$    & 39 $^{+3\phn}_{-5\phn}$
                   & \phn76 $^{+7\phn}_{-5\phn}$  & 0.44 $^{+0.03}_{-0.01}$  
                   & 2.14 $^{+0.03}_{-0.03}$      & \phd0.02 $^{+0.06}_{-0.07}$ \\
J04141458+2827580  & FN Tau            
                   & 0.044 $^{+0.002}_{-0.002}$   & $<0.32$                  & $p(\log{\alpha}); \downarrow$
                   & $>3.5$                       & $p(\gamma); \uparrow$    & $p(i)$
                   & $p(\varphi)$                 & $<0.17$                 
                   & $<1.35$                      & -1.41 $^{+0.06}_{-0.09}$ \\
J04154278+2909597  & IRAS 04125+2902   
                   & 0.042 $^{+0.001}_{-0.001}$   & 0.27 $^{+0.03}_{-0.02}$  & $>0.74$
                   & 6.8 $^{+2.4}_{-1.0}$         & -2.5 $^{+1.1}_{-0.3}$    & 26 $^{+7\phn}_{-13}$
                   & 155 $^{+41}_{-26}$           & 0.30 $^{+0.02}_{-0.01}$ 
                   & 1.68 $^{+0.02}_{-0.03}$      & -1.25 $^{+0.05}_{-0.08}$ \\
J04173372+2820468  & CY Tau            
                   & 0.207 $^{+0.012}_{-0.005}$   & 0.26 $^{+0.16}_{-0.01}$  & $p(\log{\alpha}); \uparrow$ 
                   & 4.3 $^{+3.7}_{-0.2}$         & \phd0.3 $^{+0.1}_{-2.6}$ & 27 $^{+3\phn}_{-4\phn}$
                   & 136 $^{+8\phn}_{-8\phn}$     & 0.36 $^{+0.02}_{-0.02}$  
                   & 1.67 $^{+0.03}_{-0.02}$      & -0.74 $^{+0.05}_{-0.09}$ \\
J04183444+2830302  & V410 X-ray 2      
                   & 0.026 $^{+0.002}_{-0.001}$   & 0.14 $^{+0.38}_{-0.04}$  & $p(\log{\alpha}); \downarrow$
                   & $p(\beta); \downarrow$       & $p(\gamma); \uparrow$    & 62 $^{+10}_{-27}$
                   & 106 $^{+19}_{-20}$           & 0.16 $^{+0.05}_{-0.02}$ 
                   & 1.33 $^{+0.11}_{-0.08}$      & -1.64 $^{+0.09}_{-0.09}$ \\
\enddata
\tablecomments{See notes for Table~\ref{table:SBpars}.  These measurements were originally made by \citet{tripathi17}.  They last two columns have been updated here to account for the appropriate distances inferred from {\it Gaia} DR2 parallaxes (see Table~\ref{table:starsB}).}
\end{deluxetable}
\end{longrotatetable}

\clearpage
\begin{longrotatetable}
\begin{deluxetable}{llccccc|ccc}
\tabletypesize{\scriptsize}
\tablecaption{Adopted and Inferred Stellar and Accretion Parameters\label{table:stars}}
\tablehead{
\colhead{2MASS Designation} & \colhead{Name} & \colhead{$\varpi / {\rm mas}$} & \colhead{$d / pc$} & \colhead{$\log{T_{\rm eff}/{\rm K}}$} & \colhead{$\log{L_\ast / L_\odot}$} & \colhead{notes} & \colhead{$\log{M_\ast/M_\odot}$} & \colhead{$\log{t_\ast/{\rm yr}}$} & \colhead{$\log{\dot{M}_\ast/M_\odot \, {\rm yr}^{-1}}$}
}
\colnumbers
\startdata
J15392776-3446171  & IK Lup (Sz 65) 
                                   & $6.44\pm0.05$                & $155\pm3$ 
                                   & $3.61\pm0.02$                & -$0.06\pm0.20$        & \nodata 
                                   & -0.14 $^{+0.08}_{-0.13}$ & $6.1\pm0.4$             & $<$-9.4 \\
J15392828-3446180  & Sz 66 
                                   & $6.36\pm0.09$                & $156\pm4$
                                   & $3.53\pm0.01$                & -$0.66\pm0.20$        & \nodata
                                   & -0.51 $^{+0.06}_{-0.09}$ & $6.2\pm0.3$             & \phn$-8.5\pm0.3$ \\  
J15450634-3417378  & J15450634 
                                   & $6.46\pm0.06$                & $154\pm3$ 
                                   & \nodata                            & \nodata                     & 1 
                                   & \nodata                            & \nodata                     & \nodata \\
J15450887-3417333  & J15450887 
                                   & $6.45\pm0.20$                & $154\pm6$
                                   & $3.49\pm0.01$                & -$1.20\pm0.22$        & \nodata     
                                   & -0.90 $^{+0.10}_{-0.05}$ & 6.3 $^{+0.3}_{-0.6}$ & \phn$-8.3\pm0.3$ \\
J15451741-3418283  & HW Lup (Sz 69)
                                   & $6.47\pm0.08$                & $154\pm3$ 
                                   & $3.51\pm0.01$                & -$1.03\pm0.20$       & \nodata
                                   & -0.73 $^{+0.08}_{-0.10}$ & $6.4\pm0.4$            & \phn$-9.5\pm0.3$ \\     
J15464473-3430354  & GW Lup (Sz 71) 
                                   & $6.42\pm0.06$                & $155\pm3$  
                                   & $3.56\pm0.02$                & -$0.48\pm0.20$       & \nodata
                                   & -0.34 $^{+0.10}_{-0.17}$ & $6.3\pm0.4$            & \phn$-9.0\pm0.4$ \\
\enddata
\tablecomments{Notes: (1) No spectral classification available in the literature.  (2) Anomalously low $L_\ast$.  These disks are inferred to have nearly edge-on viewing geometries (see Table~\ref{table:SBpars}), suggesting considerable self-absorption of starlight.  The mean $L_\ast$ (and standard deviation) for targets within one spectral type sub-class is adopted in the \{$M_\ast$, $t_\ast$\} inferences.  (3) Anomalously low $L_\ast$, but for unclear reasons.  See Note (2) above for mass and age calculations.  (4) The stellar parameters are especially uncertain.  For J16070545, we assign the approximate stellar parameters estimated by \citet{comeron09}.  Their $L_\ast$ value is anomalously low for the given effective temperature, but that is likely due to the nearly edge-on disk viewing geometry.  For J16070854, we have adopted the assumed M5 spectral type parameters discussed in Appendix C of \citet{alcala17}.  \citet{muzic14} suggest a considerably earlier type ($\sim$M0).  Extinction and strong veiling are problematic in this case.}
\end{deluxetable}
\end{longrotatetable}

\clearpage
\begin{longrotatetable}
\begin{deluxetable}{llccccc|ccc}
\tabletypesize{\scriptsize}
\tablecaption{Adopted and Inferred Stellar and Accretion Parameters\label{table:starsB}}
\tablehead{
\colhead{2MASS Designation} & \colhead{Name} & \colhead{$\varpi / {\rm mas}$} & \colhead{$d / pc$} & \colhead{$\log{T_{\rm eff}/{\rm K}}$} & \colhead{$\log{L_\ast / L_\odot}$} & \colhead{notes} & \colhead{$\log{M_\ast/M_\odot}$} & \colhead{$\log{t_\ast/{\rm yr}}$} & \colhead{$\log{\dot{M}_\ast/M_\odot \, {\rm yr}^{-1}}$}
}
\colnumbers
\startdata
J03454828+3224118  & LkH$\alpha$ 330   
                   & $3.22\pm0.08$               & $309\pm13$
                   & $3.77\pm0.03$     & \phd$1.36\pm0.20$ & 1, 2
                   & \phd0.47 $^{+0.06}_{-0.09}$ & $6.1\pm0.4$    & \phn-$7.6\pm0.5$ \\
J04141458+2827580  & FN Tau
                   & $7.62\pm0.06$               & $131\pm2$ 
                   & $3.49\pm0.02$     & -$0.14\pm0.10$    & 3
                   & -0.67 $^{+0.11}_{-0.04}$    & $<5.7$                     & \nodata \\
J04154278+2909597  & IRAS 04125+2902   
                   & $6.25\pm0.07$               & $159\pm3$ 
                   & $3.56\pm0.02$    & -$0.39\pm0.06$    & 3, 4 
                   & -0.34 $^{+0.10}_{-0.10}$    & $6.2\pm0.3$     & \phn-$9.4\pm0.5$ \\
J04173372+2820468  & CY Tau            
                   & $7.76\pm0.04$               & $129\pm2$  
                   & $3.56\pm0.02$    & -$0.48\pm0.10$    & 3, 5, 6
                   & -0.35 $^{+0.11}_{-0.12}$    & $6.3\pm0.3$     & \phn-$8.4\pm0.5$  \\
J04183444+2830302  & V410 X-ray 2      
                   & $7.75\pm0.48$               & 129 $^{+10}_{-7\phn}$ 
                   & $3.59\pm0.02$    & \phd$0.35\pm0.11$ & 3 
                   & -0.32 $^{+0.09}_{-0.06}$    & $<5.7$                     & \nodata \\
J04245708+2711565  & IP Tau            
                   & $7.66\pm0.04$               & $130\pm2$
                   & $3.59\pm0.02$    & -$0.39\pm0.13$     & 3, 7
                   & -0.23 $^{+0.13}_{-0.13}$    & $6.4\pm0.3$    & \phn-$8.3\pm0.5$ \\
J04295156+2606448  & IQ Tau            
                   & $7.62\pm0.06$               & $131\pm2$
                   & $3.58\pm0.02$    & -$0.15\pm0.12$     & 3, 8, 9
                   & -0.30 $^{+0.11}_{-0.14}$    & $5.9\pm0.3$   & \phn-$7.9\pm0.5$ \\
\enddata
\tablecomments{The initial $T_{\rm eff}$ and $L_\ast$ estimates were derived from the first listed reference code in the `notes' column (Col.~7).  The initial accretion luminosities were derived from the subsequent reference codes (if more than one, the average of the two quoted references was adopted).  If no uncertainties on these parameters were specified in these references, we presumed log-normal standard deviations of 0.02, 0.2, and 0.5\,dex for $\log{T_{\rm eff}/{\rm K}}$, $\log{L_\ast/L_\odot}$, and $\log{L_{\rm acc}/L_\odot}$, respectively.  The reference codes correspond to: 1 = \citet{andrews11}, 2 = \citet{manara14}, 3 = \citet{andrews13}, 4 = \citet{espaillat15}, 5 = \citet{gullbring98}, 6 = \citet{white01}, 7 = \citet{ingleby13}, 8 = \citet{valenti93}, 9 = \citet{hartmann98}, 10 = \citet{espaillat10}, 11 = \citet{muzerolle98}, 12 = \citet{herczeg08}, 13 = \citet{white04}, 14 = \citet{calvet04}, 15 = \citet{donehew11}, 16 = \citet{natta04}, 17 = \citet{garcia-lopez06}, 18 = \citet{mendigutia11}, 19 = \citet{andrews12}, 20 = \citet{fairlamb15}, 21 = \citet{barenfeld16}, 22 = \citet{mathews12}, 23 = \citet{andrews10b}, 24 = \citet{natta06}, 25 = \citet{andrews09}, 26 = \citet{cieza12}, 27 = \citet{eisner05}, 28 = \citet{salyk13}.}
\end{deluxetable}
\end{longrotatetable}

\section{Host Population Model} \label{appendixB}

To discuss the relationships between disk and host star properties in the general population, we first need to design a model of the stellar properties in our joint sample.  The goal is to build a framework from which random draws would produce a sample that is statistically similar to the multi-dimensional host star properties derived here (in Tables~\ref{table:stars} and \ref{table:starsB}); in effect, a model that can reproduce the behavior in Figure~\ref{fig:stars_covar}.  In principle, given the inter-relationships between these properties one could pick any two-dimensional projection of that parameter-space (any panel in Figure~\ref{fig:stars_covar}), generate a parametric description of its behavior, and then naturally reproduce the other projections (panels). But in practice, it makes sense to quantify the population in a projection that can be parameterized relatively easily.

\begin{figure}[t!]
\includegraphics[width=\linewidth]{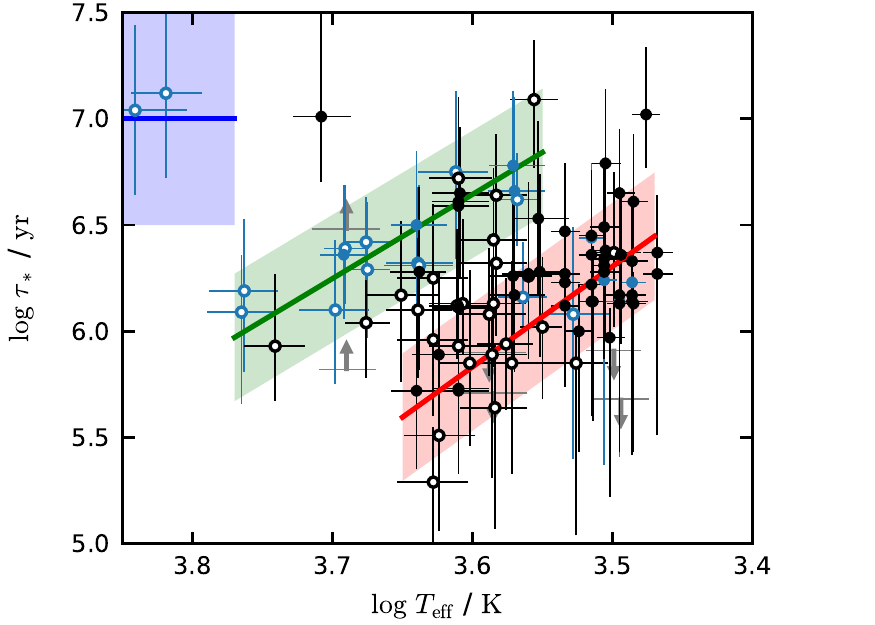}
\figcaption{An illustration of the underlying construction of the host star population model, showing a three-part linear model of (visual, not necessarily physical) sub-populations {\tt a} (red), {\tt b} (green), and {\tt c} (blue).  The model component mean values are shown as darker lines, and their corresponding (Gaussian) dispersions are marked as lighter shaded regions.  The datapoints are as shown in the lower left corner of Figure~\ref{fig:stars_covar}.  
\label{fig:tripartite}
}
\end{figure}

After some experimentation, we decided to manually tune (i.e., adjust by visual experimentation) a three-part linear model in the \{$\log{T_{\rm eff}}$, $\log{\tau_\ast}$\} plane.  Figure~\ref{fig:tripartite} directly illustrates this model behavior.  The three sub-populations of interest were identified manually; they are technically arbitrary, but are suitable for achieving our goals.  Each sub-population is characterized with a linear mean model and a Gaussian dispersion in $\log{\tau_\ast}$, such that samples would be drawn from
\begin{equation}
    \log{\tau_\ast} = \mathcal{A} + \mathcal{B} \, \log{T_{\rm eff}} + \mathcal{N}(0, \sigma),
    \label{eq:agemodel}
\end{equation}
where $\mathcal{N}(0, \sigma)$ indicates a random Gaussian deviate with mean zero and standard deviation $\sigma$.  For the cooler, younger sub-population {\tt a} (red in Figure~\ref{fig:tripartite}), the targets are described well with $\mathcal{A}_{\tt a} = 22.9$, $\mathcal{B}_{\tt a} = -4.7$, and $\sigma_{\tt a} = 0.3$ on the interval $\log{T_{\rm eff, {\tt a}}} \in [3.47, 3.65]$.  At overlapping, but slightly warmer and older temperatures and ages, respectively, sub-population {\tt b} (green in Figure~\ref{fig:tripartite}) is characterized by $\mathcal{A}_{\tt b} = 20.9$, $\mathcal{B}_{\tt b} = -4.0$, and $\sigma_{\tt b} = 0.3$ on the interval $\log{T_{\rm eff, {\tt b}}} \in [3.55, 3.77]$.  The few remaining sample targets include the early type stars in sub-population {\tt c} (blue in Figure~\ref{fig:tripartite}), for which we set a constant mean age with a broad distribution: $\mathcal{A}_{\tt c} = 7.0$, $\mathcal{B}_{\tt b} = 0$, and $\sigma_{\tt b} = 0.5$ for $\log{T_{\rm eff, {\tt c}}} \in [3.77, 4.00]$.  A model host star population of size $N$ is then created as follows: (1) evenly distribute 0.72\,$N$ points over the interval $\log{T_{\rm eff, {\tt a}}}$, 0.25\,$N$ points over $\log{T_{\rm eff, {\tt b}}}$, and 0.03\,$N$ points over $\log{T_{\rm eff, {\tt c}}}$; (2) for each sub-population, compute $\log{\tau_\ast}$ values according to Eq.~\ref{eq:agemodel}.  

The fact that we have broken down the behavior into sub-populations is not to imply that these are necessarily {\it physically} distinct.  The uncertainties on individual host parameters are typically large enough to be consistent with a single age (although see Section~\ref{subsec:trans}) -- it just represents a convenient way to reproduce the inferred parameters for the full population.  
.

\end{document}